\documentstyle{l-aa} %Astronomy & Astrophysics macro
\input{epsfig.sty}

\begin{document}

% ****** definitions used below **********
\def\Mo{M_\odot}
% ********************************************

   \thesaurus{04  % A&A Section 4: Galactic structures and dynamics
              (11.16.1;  % Galaxies: photometry
               11.19.2;  % Galaxies: spiral
%              11.19.7;  % Galaxies: statistics
               11.06.2;  % Galaxies: fundamental parameters
               11.19.6)} % Galaxies: structure

   \title{Near-IR photometry of disk galaxies:
          search for nuclear isophotal twist and double bars}

%   \subtitle{}

   \author{Bruno Jungwiert \inst{1,2}, Fran\c coise Combes \inst{1}
           and Dave J. Axon \inst{3}}

   \offprints{B. Jungwiert \inst{1}}

   \institute{\vskip3mm
\inst{1} DEMIRM, Observatoire de Paris, 61 Avenue de l'Observatoire,
75 014 Paris, France\\
\inst{2} Astronomical Institute, Academy of Sciences of the
Czech Republic,
Bo\v{c}n\'{\i} II 1401, 141 31 Prague 4, Czech Republic\\
\inst{3} ESA secondment, Space Telescope Science Institute, Baltimore,
MD 21218, USA \\
%e-mail: bruno@astro.ig.cas.cz \\
%\inst{4} Universit\'e Paris VII \& Observatoire de Meudon, 5 Place Jules
%Janssen, 92 190 Meudon, France, e-mail: bruno@mesiom.obspm.fr
}

   \date{Received 29 November 1996 / Accepted 21 January 1997}

   \maketitle
\markboth{B. Jungwiert et al.: Isophotal twist in disk galaxies}
{}

\begin{abstract}
We present a near-IR, mainly H band, photometry of 72 nearby
($d < 40$ Mpc) disk galaxies. The main goal of the survey was
to search for isophotal twist inside their nuclear regions.
As the twist can be due in some cases to projection effects,
rather than resulting from a dynamical phenomenon, we
deproject -- under the simplifying assumption of a 2D geometry --
all galaxies whose disk position angle and inclination are known,
the latter not exceeding $75^o$. We show the ellipticity,
position angle and surface brightness radial profiles,
and discuss how a projection of 2D and 3D bars can distort
the isophotes, give an illusion of a non-existing double bar
or mask a real one.
We report 15 new double-barred galaxies and confirm 2
detected previously. We identify 14 additional twists
not known before and we also find nuclear triaxial structures in
three SA galaxies. The frequency of Seyferts among galaxies
with nuclear bars or twists is high.
Since these observations are part of a larger survey,
the interpretation of the results will be given in a future paper,
as soon as the number of objects grows enough to permit meaningful
statistics.
As a secondary product, we publish structural parameters
(length and axis ratio) of large-scale bars in order to extend
still scarce data on bars in the near-IR.

%\vskip-1mm
      \keywords{galaxies: photometry -- galaxies: spiral, structure of --
galaxies: fundamental parameters}
\end{abstract}

\section{Introduction}

Non-axisymmetric distortions (bars, ovals) in inner parts of disk
galaxies are recognized, due to N-body simulations including gas
(e.g. Friedli \& Benz 1993, Combes 1994), to be an
efficient mechanism for driving the interstellar medium (ISM) into
the nuclear region. Numerous observations complete the picture by
showing that various kinds of central activity, like Seyfert nuclei
(e.g. Hummel et al. 1987), LINERs and starbursts
(e.g. Devereux 1989, Telesco et al. 1993) are often correlated
with the presence of bars
(note however counter-examples
of many Seyferts (McLeod \& Rieke 1995), as well as galaxies with strong IR
excess (Pompea \& Rieke 1990), that do not show any non-axisymmetric
deviation).

While a large-scale bar is probably sufficient to fuel a
starburst occuring inside a few hundred parsecs,
it seems unable to bring the ISM down to the scale governed
by a central blackhole ($<10$ pc) that presumably powers active
galactic nuclei
(AGNs). The ``bar-within-bar'' scenario was proposed
by Shlosman et al. (1989) to overcome the problem:
first, a large-scale bar accumulates gas in a sub-kpc nuclear
disk, that undergoes, when becoming massive enough, a secondary
bar-forming instability, susceptible to funnel the ISM down
to the BH region. The possibility to create
such a double-bar, with the inner component rotating at a higher
angular rate, was demonstrated in simulations of Friedli \&
Martinet (1993).

In turn, searches for inner isophotal twists that would
observationally confirm the existence of double-bar configurations
were initiated. Preexisting detections of twists in spiral
galaxies (de Vaucouleurs 1974; Kormendy 1979,1982; Jarvis et al. 1988;
Pompea \& Rieke 1990; Buta 1990; Buta \& Crocker 1993, BC93 hereafter) were
substantially multiplied due to near-IR observations of
Shaw et al. (1993, 1995) and
BVRI survey of Wozniak et al. (1995, W95 hereafter). Further twists
in the near-IR were reported
by Elmegreen et al. (1996, E96 hereafter) who summarized the
preceding surveys: 51 isophote twists were discovered amongst
80 barred spirals and lenticulars (the frequency is
insignificant since galaxies where the twist was
expected were observed preferentially). The
last authors also examined blue plates
in the Sandage \& Bedke atlases and found 18 additional galaxies
displaying the twist. Finally, Rauscher (1995) adds 5 other examples
in the near-IR, raising the number of twist detections to 74.

A (considerable) part of twists can be due to projection effects
on triaxial structures -- bars, bulges or combination of both
-- with varying excentricity but no intrinsic (i.e. face-on viewed)
variation of isophotal position angle.

For the intrinsic twists, a scenario competitive with
the bar-within-bar hypothesis was suggested by
Shaw et al. (1993) on the grounds of the orbital structure
inside a bar: a gaseous ring between two inner Lindblad
resonances (ILRs), phase-shifted with respect to the main bar
due to its association with $x_2$ orbits, perturbs gravitationally
the stellar component thus causing its isophote twist. In contrast
to the double-bar of Friedli \& Martinet, the perturbed region can be
tilted only towards the leading side of the main bar,
both components rotating at the same pattern speed.

Recently, Davis \& Hunter (1995) and Friedli (1996) extended the
panorama of double-bar dynamics by considering counter-rotating nuclear
bars.

There are numerous open questions concerning the twists in disk galaxies:
What is the frequency of the twist phenomenon~? What is the fraction
of intrinsic twists~? How frequent are triaxial bulges~?
Is there a significant correlation between the intrinsic twists
and the presence of nuclear activity~? Are the intrinsic twists
correlated with inner/nuclear rings~? What mechanism is responsible
for the intrinsic twists: bar-within-bar instability or gas perturbing
stars between the ILRs, or both~? How the twist properties vary along
the Hubble sequence~?

The above questions have no definitive answers mainly because
of incompleteness of existing surveys, their bias towards galaxies
with enhanced nuclear activity, insufficient resolution close
to galactic centers where the twists occur, and projection effects.

This survey is intended to enlarge the set of disk galaxies showing
the nuclear isophote twist and to quantify it for future statistical
purposes. The paper is organized as follows: Section 2 summarizes
observations and data reduction procedures including the ellipse
fitting on isophotes. Section 3 deals, on a qualitative level,
with projection and deprojection of bars and double bars, since this
problem is crucial for establishing meaningful statistics and
conclusions about the nature of twists. Individual galaxies are
shortly described in Section 4, conclusions are outlined in Section 5.
The contour plots as well as profiles of ellipticity,
position angle and surface brightness
along bars are given in the appendix.

\section{Observations and data reduction}

The data (Table 1) were acquired on three nights
(February 27 to March 1, 1995)
using the infrared camera IRAC2 installed on the 2.2-meter telescope
at the ESO's La Silla Observatory. This camera is equipped with a
Hg:Cd:Te NICMOS3 array of 256 x 256 pixels. The detector scale
was chosen to be 0.52 arcsec/pixel corresponding to the field
of view of about 2 x 2 arcminutes.
The seeing on the first night was 1.2'' (FWHM) for all the three
filters; during the second and the third nights it got reduced to 0.9''
and 1.0'' in the H-band (observations in bands K and J were
carried out during the first night only).

%------------------------------------------------------------------
\begin{table}
\caption{Observed galaxies} \label{tbl-1}
\vskip-5mm
\begin{center}\scriptsize
\begin{tabular}{llrrrl}
Galaxy & Type  & $\tau_H$ & $D_{25}$ & d     & Nuclear activity,\\
       & (RC3) & (sec)        & ('') & [Mpc] & nucl. rings (nr), \\
       &       &              &      &       & nucl. spirals (ns)\\
%\multicolumn{1}{c}{$P$\tablenotemark{a}} & $P R_{maj}$ & $P R_{min}$ &
%\multicolumn{1}{c}{$\Theta$\tablenotemark{b}} \\
%\tableline
     &    \\
N 613  & SB(rs)bc    & 3x50 & 380     & 17.9 & Sey$^{a}$, ns$^d$ \\
N 1079 & RSAB(rs)0/a & 4x50 & 208     & 17.1 & \\
N 1187 & SB(r)c      & 4x50 & 330     & 16.7 & \\
N 1255 & SAB(rs)bc   & 4x50 & 250     & 20.3 & \\
N 1302 & RSB(r)0/a   & 4x50 & 233     & 20.5 & nr$^d$\\
%N 1313 & SB(s)d     & 4x50 & 547     &      & \\
N 1353 & SB(rs)b     & 3x50 & 203     & 18.4 & \\
N 1365 & SB(s)b      & 2x50 & 673     & 19.4 & Sey 1/H~II$^{a,b}$, ns$^d$\\
N 1398 & R'SB(r)ab   & 2x50 & 425     & 16.5 & \\
%N 1404 & E1         & 2x50 & 199     & 23.3 & \\
N 1433 & R'SB(r)ab   & 4x30 & 387     & 11.1 & nr$^d$\\
%N 1510 & S0?   & 4x50 & 79     & 10.0 & \\
N 1512 & SB(r)a      & 4x50 & 535     & 11.1 & nr$^d$\\
N 1518 & SB(s)dm     & 4x50 & 181     &      & \\
N 1640 & SB(r)b      & 3x50 & 158     & 19.2 & \\
N 1744 & SB(s)d      & 4x50 & 488     &  7.4 & \\
N 1784 & SB(r)c      & 4x50 & 239     & 29.2 & \\
N 1792 & SA(rs)bc    & 4x50 & 315     & 13.2 & \\
N 1808 & RSAB(rs)a   & 3x50 & 387     & 10.4 & H~II$^{a,b}$, nr'$^d$\\
N 1832 & SB(r)bc     & 4x50 & 154     & 24.0 & \\
N 2217 & RSB(rs)0+   & 4x50 & 268     & 19.0 & \\
N 2442 & SAB(s)bc    & 2x50 & 330     & 15.5 & \\
N 2525 & SB(s)c      & 4x50 & 173     & 19.3 & \\
N 2811 & SB(rs)a     & 4x50 & 151     & 31.6 & \\
N 2911 & SA(s)0      & 4x50 & 244     & 42.3 & Sey 3$^a$\\
N 2935 & R'SAB(s)b   & 4x50 & 218     & 28.3 & nr$^d$\\
N 2997 & SAB(rs)c    & 3x50 & 535     & 11.9 & nr$^d$\\
%N 3080 & Sa         & 1x50 &     & 142. & Sey 1$^a$\\
N 3166 & SAB(rs)0/a  & 4x50 & 287     & 17.4 & \\
N 3346 & SB(rs)cd    & 4x50 & 173     & 17.1 & \\
N 3368 & SAB(rs)ab   & 2x150& 455     & 12.2 & \\
N 3384 & SB(rs)0-    & 4x50 & 330     & 10.1 & ?$^a$\\
N 3393 & R'SB(rs)a   & 4x50 & 131     & 47.6 & Sey 2$^a$\\
N 3593 & SA(s)0/a    & 4x50 & 315     &  8.7 & nr$^d$\\
N 3637 & RSB(r)0/a   & 4x50 &  95     & 23.6 & \\
N 3673 & SB(rs)b     & 4x50 & 218     & 23.9 & \\
N 3885 & SA(s)0/a    & 4x50 & 144     & 22.1 & \\
N 3887 & SB(r)bc     & 4x50 & 199     & 14.8 & \\
N 4050 & SB(r)ab     & 4x50 & 185     & 23.6 & \\
N 4106 & SB(s)0+     & 4x50 &  97     & 26.8 & \\
N 4178 & SB(rs)dm    & 4x50 & 308     &      & \\
N 4192 & SAB(s)ab    & 4x50 & 586     &      & Sey 3$^a$\\
N 4212 & SAc         & 4x50 & 190     &      & \\
N 4216 & SAB(s)b     & 4x50 & 488     &      & \\
N 4267 & SB(s)0-     & 4x50 & 194     & 14.8 & \\
%N 4273 & SBc   & 4x50 & 141     & 31.9 & H~II (D89)\\
N 4424 & SB(s)a      & 4x50 & 218     &      & \\
N 4438 & SA(s)0/a    & 4x50 & 511     &      & Sey 3$^a$\\
N 4442 & SB(s)0      & 4x50 & 274     &  7.7 & \\
N 4454 & RSB(r)0/a   & 4x50 & 120     & 30.3 & \\
N 4461 & SB(s)0+     & 4x50 & 213     & 26.4 & \\
N 4501 & SA(rs)b     & 4x50 & 415     & 31.2 & Sey 2$^a$\\
N 4503 & SB0-        & 4x50 & 213     & 18.7 & \\
N 4519 & SB(rs)d     & 2x50 & 190     & 16.7 & \\
N 4546 & SB(s)0-     & 2x50 & 199     & 13.7 & \\
N 4612 & SB(r)0+     & 2x50 & 239     & 24.8 & \\
N 4665 & SB(s)0/a    & 4x50 & 228     & 10.6 & \\
N 4684 & SB(r)0+     & 3x50 & 173     & 20.8 & \\
N 4689 & SA(rs)bc    & 4x50 & 256     & 22.4 & \\
N 4694 & SB0         & 3x50 & 190     & 16.4 & H~II$^a$\\
N 4731 & SB(s)cd     & 4x50 & 396     & 19.5 & \\
N 4781 & SB(rs)d     & 4x50 & 208     & 16.1 & \\
N 4856 & SB(s)0/a    & 3x50 & 256     & 17.0 & \\
N 4900 & SB(rs)c     & 3x50 & 134     & 13.1 & \\
N 4902 & SB(r)b      & 4x50 & 181     & 35.2 & \\
N 4984 & RSAB(rs)0+  & 4x50 & 165     & 15.2 & ?$^c$, nr$^d$\\
N 5101 & RSB(rs)0/a  & 4x50 & 322     & 23.2 & \\
N 5236 & SAB(s)c     & 4x50 & 773     &      & H~II$^b$, nr$^d$\\
N 5427 & SA(s)c      & 4x50 & 169     & 35.3 & Sey 2$^a$, nr$^d$\\
N 5566 & SB(r)ab     & 4x50 & 396     & 20.8 & \\
N 5643 & SAB(rs)c    & 4x50 & 274     & 13.7 & Sey 2$^a$\\
N 5701 & RSB(rs)0/a  & 3x50 & 256     & 20.9 & \\
N 6753 & RSA(r)b     & 4x50 & 147     & 39.3 & nr$^d$\\
N 6782 & RSAB(r)a    & 4x50 & 131     & 48.5 & nr$^d$\\
N 6810 & SA(s)ab     & 1x50 & 190     & 23.5 & \\
E437-67 & R'SB(r)ab  & 4x50 & 123     & 39.4 & nr$^d$\\
I 1953 & SB(rs)d     & 3x50 & 165     & 22.9 & \\
\end{tabular}
\end{center}
\end{table}

Typically (but not always; see Table 1),
four object frames were obtained for a galaxy
in one band: exposure length for filters H, J and K
was respectively 50s (achieved by 5 elementary integrations
of 10s each, in order to avoid the detector saturation),
30s (1x30s) and 50s (10x5s), resulting in the total
integration time of 200s, 120s and 200s.
To reduce the contamination by defective pixels (less than 1\%),
the telescope pointing was shifted by a few arcseconds for every
object frame.

Since the sky in the near-IR varies on the timescale
of the total integration time, a sky frame (of the same exposure
length as for an object frame) was taken after each object frame:
the typical observing sequence was thus OBJECT-SKY-O-S-O-S-O-S.
The sky frames were offset from a galaxy by a few arcminutes.
Dark current frames of all relevant exposure times were
prepared as well.

The data was reduced by means of the ESO MIDAS package. First,
from each object frame the subsequent sky frame was subtracted (no
dark subtraction was needed here because of equal exposure lengths).
The resulting images were divided by the flatfield (normalized to unity)
to  eliminate the variation in the pixel-to-pixel response
(about 10\%); the flatfield frame was constructed for each galaxy
separately by median combining of dark-subtracted sky frames.
In turn, the sky-subtracted and flatfielded images were aligned
and averaged into one frame that was cleaned from remaining bad pixels
(bi-linear interpolation) and intervening stars
(bi-quadratic interpolation).

%table caption
Table (1) summarizes basic information about observed objects:
{\it Col.(1)} Galaxy identification (N=NGC, E=ESO, I=IC), {\it Col.(2)}
 Type according
to RC3 (de Vaucouleurs et al. 1993), {\it Col.(3)} Exposure time in filter H;
four galaxies were observed also in K: NGC 1433 (4x30s), 3346 (2x50),
3887 (4x50), 5236 (4x50), and five in J: NGC 1433, 3384, 3593, 3887, 5236
(4x30s all),
{\it Col.(4)} 25 B-mag/arcsec$^2$ isophotal diameter (from the Lyon-Meudon
Extragalactic Database (LEDA), Paturel et al. 1989), {\it Col.(5)}
 kinematical
distance corrected for the Virgocentric inflow, H$_o$=75 km/s/Mpc
(from LEDA), {\it Col.(6)} Nuclear activity, rings, spirals: from
(a) V\'eron-Cetty \& V\'eron (1996), (b) Telesco et al (1993),
(c) Devereux (1989), (d) Buta \& Crocker (1993).
%end of table caption

%----------------------------------------------------------------------

%\command \subsection causes error !
%\subsection{Calibration}

\vskip3mm
{\noindent \it 2.1 Calibration}
\vskip3mm

To calibrate images, three standard
infrared stars were observed each night. The rms error
in the determination of the photometric zero points was 0.03 mag
for all three filters on the first night. The zero points for the
second and third night in band H were consistent to within the
error with that for the first night and all the three
were averaged to give the single zero point.
The airmass correction was applied using the mean atmospheric
extinction coefficients for the observing site: $a_H=0.06, a_J=0.08,
a_K = 0.11$ mag/airmass; the airmass falls between 1 and 2.03 for our
observations.

To test the photometric reliability, we have compared the results
of our calibration to published photometry.
In the H band, our sample has nine
galaxies in common (NGC 1302, 1398, 1433, 1808, 2217, 5566, 5701,
6753 and 6810) with the aperture photometry of Griersmith et
al. (1982). We have simulated the apertures of diameter 22'', 33''
and 56'' on our frames and found a mean magnitude difference,
$\Delta m_H = m_{H,ours} - m_{H,G82}$, of  $-0.19\ (\pm  0.11)$,
$-0.14\ (\pm  0.10)$ and $-0.14\ (\pm 0.09)$. The H-band aperture photometry
of 10 other galaxies of our survey (NGC 3166, 3885, 3887, 4212,
4273, 4501, 4781, 4900, 4902 and 4984) was done by Devereux (1989):
our magnitudes for his 9.3'' aperture differ by $\Delta m_H =
-0.09\ (\pm 0.12)$.
Both comparisons given above could
indicate a systematic offset of our calibration by 0.1-0.2 mag, however
this number is within the errors quoted in the referenced papers.
Another galaxy (NGC 2997) was measured by Forbes et al. (1992): in this
case $\Delta m_H = + 0.05$ and $+ 0.07$ for the 6'' and 12'' apertures.
Finally, the surface photometry of H\'eraudeau et al. (1996)
has one common object with us, NGC 6810, for which we find $\Delta m_H =
+0.05\ (\pm 0.08)$ along the 60'' major-axis profile.

%\command \subsection causes error !
%\subsection{Ellipse fitting}
\vskip3mm
{\noindent \it 2.2. Ellipse fitting}
\vskip3mm

To follow the isophotal twist, we have used the ellipse fitting
algorithm FIT/ELL3 (in the MIDAS context SURFPHOT), developed
by Bender \& M\"ollenhoff (1987) for the study of the isophotal
twist in elliptical galaxies (cf. Sect. 3). To parametrize the bars
and double bars we use terms and quantities introduced by W95
to whom we refer the reader for details:
typically, for a nearly face-on galaxy
(projection effects are discussed in Sect. 3)
with two bars, the {\it ellipticity} ($e=1-b/a$, where $a$ and $b$
are the ellipse semi-major and semi-minor axes)
first grows to a first local maximum
$e_{max}^s$ (at $a=l_{max}^s$)
corresponding to the {\it secondary} (i.e. inner) {\it bar},
then falls to a minimum $e_{min}^s$ before climbing again to a
{\it primary bar} maximum, $e_{max}^p$ (at $a=l_{max}^p$),
after which it decreases towards the
ellipticity of the disk, $e_{disk}$ (see Fig.~1 in W95). We define
the {\it sizes} of the bars by $l_{max}^s$ and $l_{max}^p$.
{\it Position angles} (measured from the North counterclockwise) of the bars
and the disk are denoted $PA^s$, $PA^p$ and $PA_{disk}$. When the $PA$
changes along a bar, we define $PA^s$ and $PA^p$ to be the $PA$
at $l_{max}^s$ and $l_{max}^p$, respectively. In agreement with
W95 and E96, we shall classify the bar isophotes
as twisted whenever the variation of the $PA$ along a bar exceeds
$10^o$.

In the appendix, we present for individual galaxies the PAs and
ellipticities plotted against the semi-major axis of the fitted
ellipses which is scaled logarithmically in order to better see
inner regions. We do not comment on any feature inside $a=3''$
since the ellipse fitting on artificial bars of known shapes
proved not to be reliable there due to the seeing
and small number of pixels. However, we show the profiles down
to $a=1''$ since they often display a continuity below $a=3''$
and might provide a reference for eventual future observations
with higher resolution. The unreliable region, $a<3''$, is
separated by a vertical dash-dot line in plots.

\section{Projection effects on isophotes: twists in ellipticals and bars}

%                                     One column figure
%_____________________________________1
\begin{figure} [htbp]
%  \picplace{8cm}
%  \begin{center}
%\leavevmode
\epsfig{file=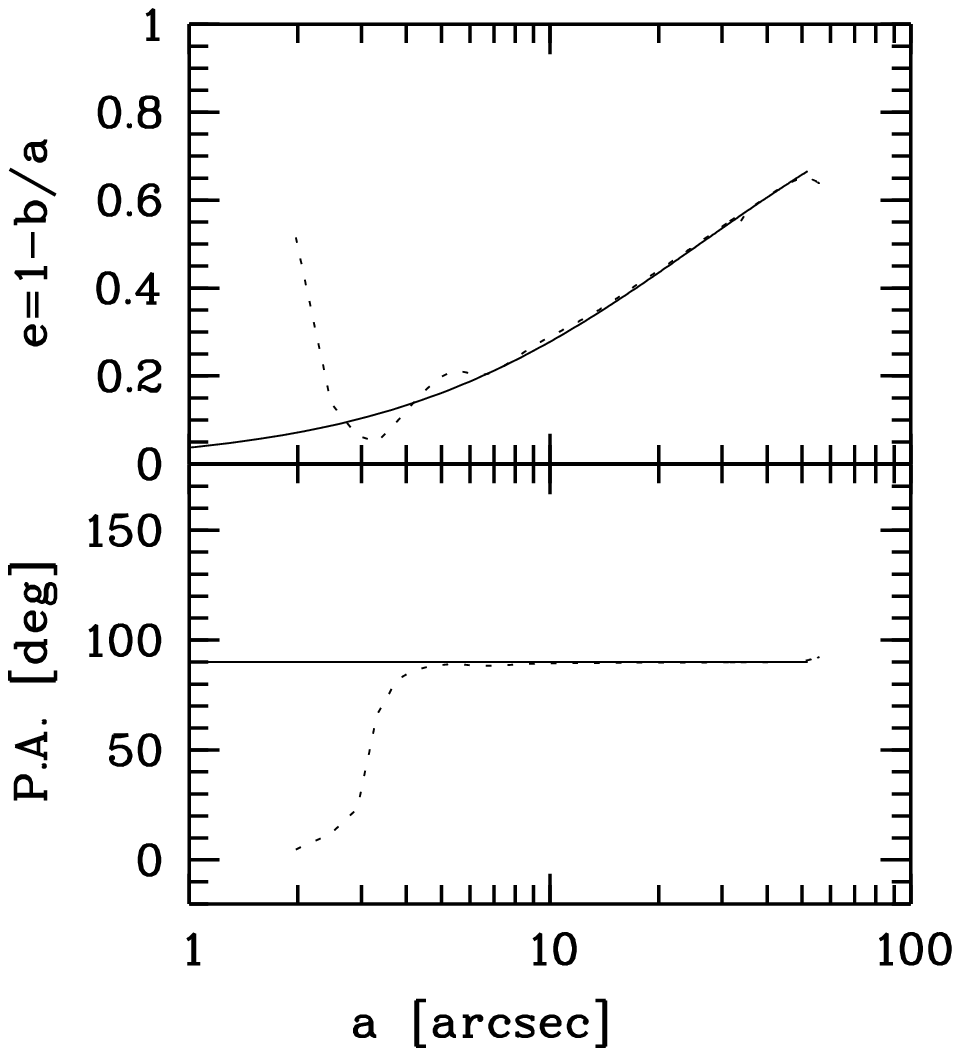, height=6.5cm, width=10cm, angle=-90}
\vskip-6.8cm
{\hskip+4cm
\epsfig{file=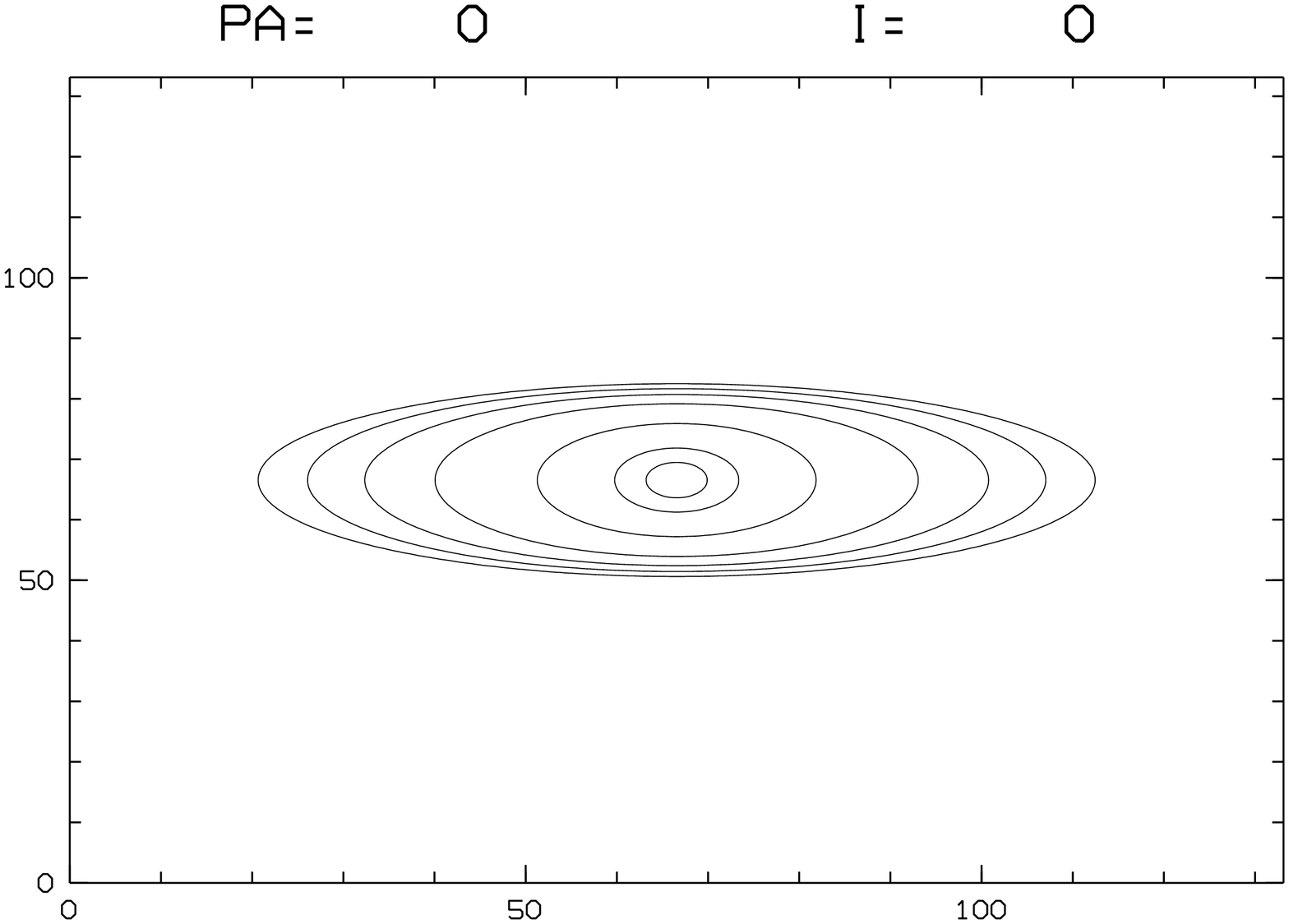, height=4.54cm, width=4.36cm, angle=-90}}
\vskip3mm
\epsfig{file=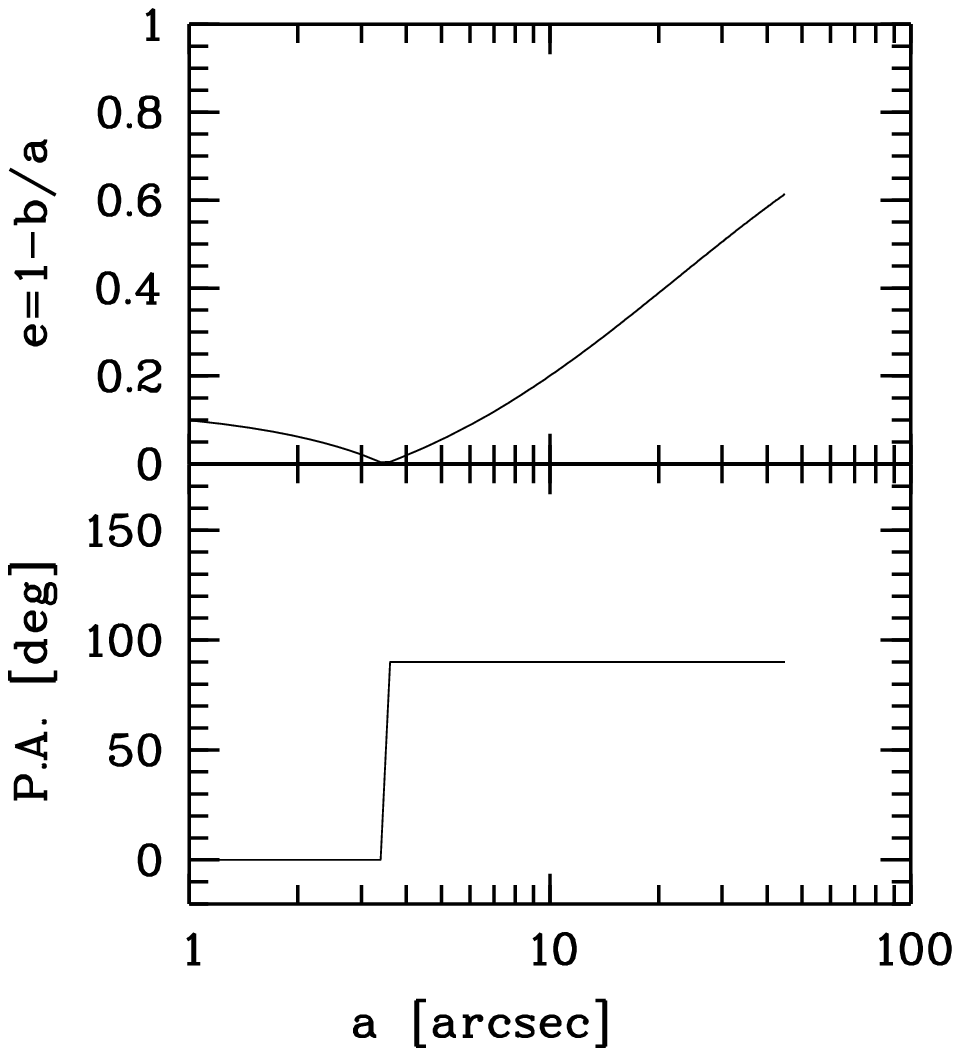, height=6.5cm, width=10cm, angle=-90}
\vskip-6.8cm
{\hskip+4cm
\epsfig{file=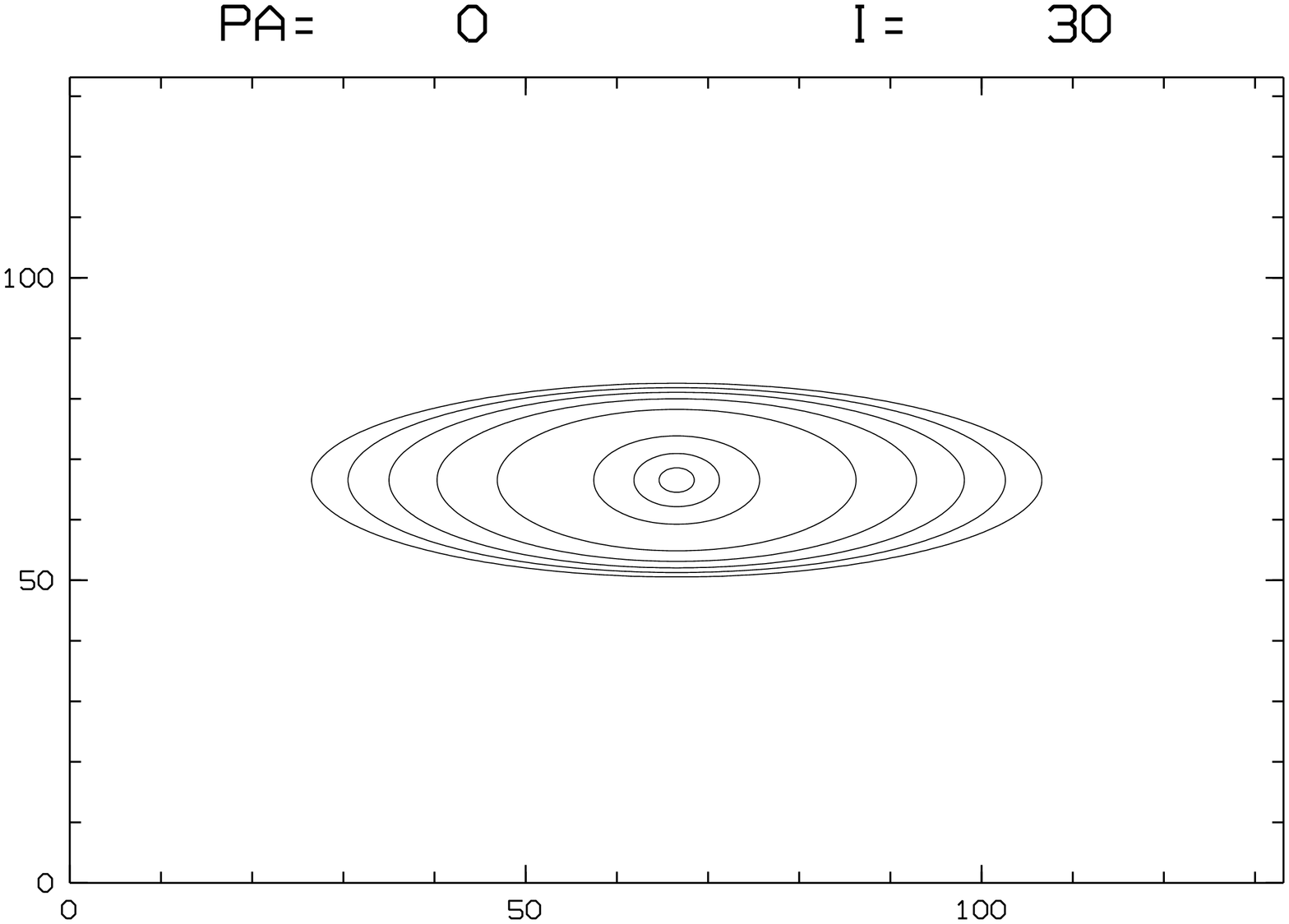, height=4.54cm, width=4.36cm, angle=-90}}
\vskip3mm
\epsfig{file=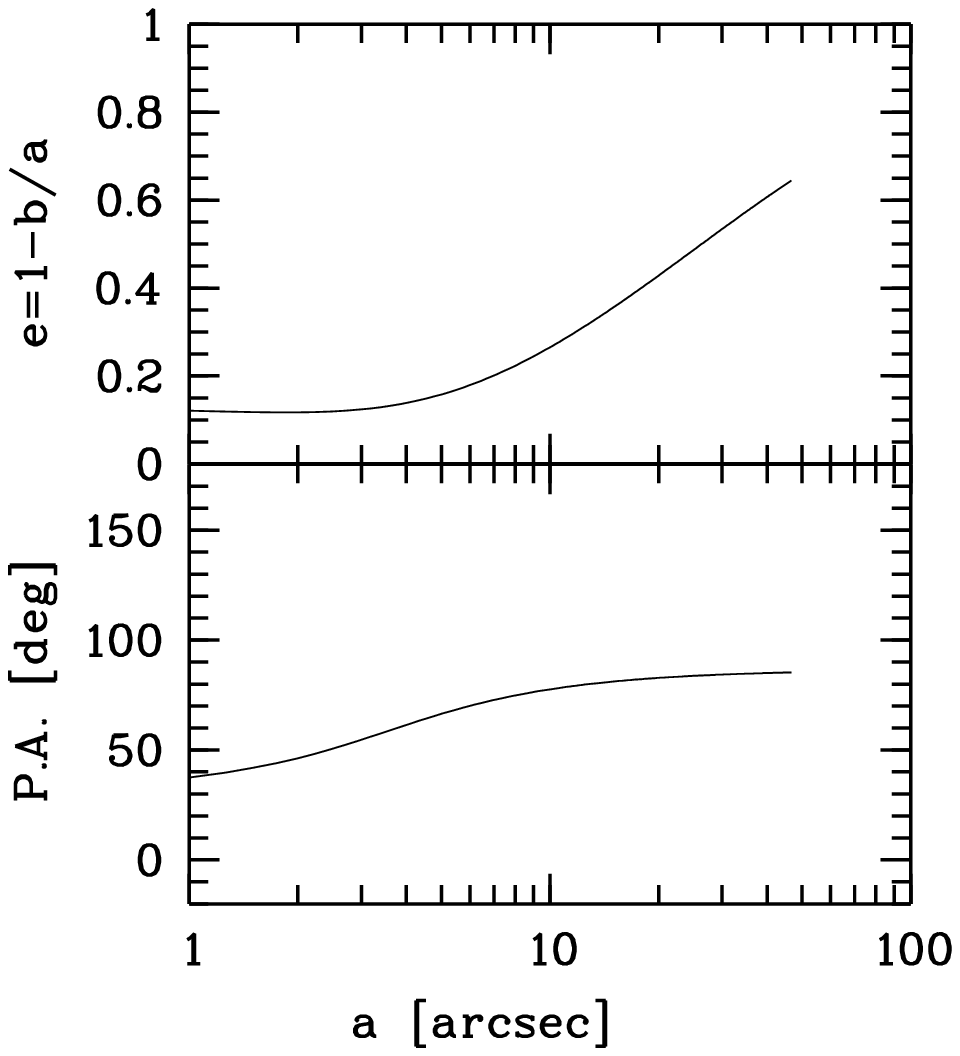, height=6.5cm, width=10cm, angle=-90}
\vskip-6.8cm
{\hskip+4cm
\epsfig{file=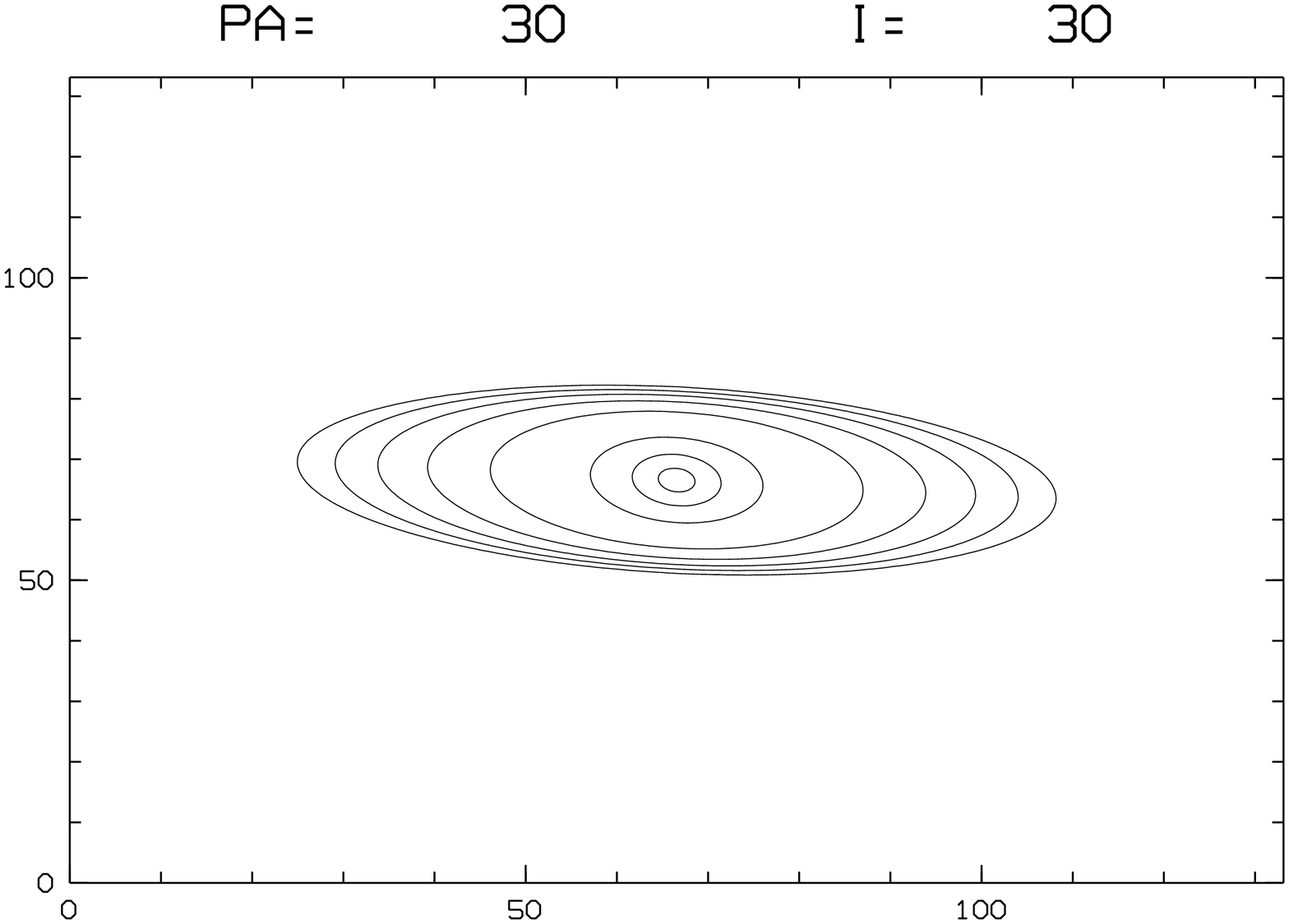, height=4.54cm, width=4.36cm, angle=-90}}
     \caption[]{Ellipticities (1-b/a), position angles (PA) and
contour plots for a single 2D bar: {\bf a)} Face-on view -- full lines in
plots of radial profiles (dotted lines indicate the same quantities
measured after the bar is first analytically projected and then
numerically deprojected back; see the text),
{\bf b)} Projection with $I=30^o$ and $PA_{proj}=0^o$,
{\bf c)} Projection with $I=30^o$ and $PA_{proj}=30^o$
                }
%  \end{center}
\end{figure}
%--------------------------------------------------------------

It has been known for more than three decades that axes of isophotal
contours in many {\it elliptical} galaxies rotate (e.g. Liller 1960,
Bertola \& Galletta 1979, Nieto et al. 1992).
Such twists can be explained either by intrinsic misalignement
of isophotal surfaces (which are ellipsoidal in the first approximation)
or by projection effects: in the latter case, the aligned ellipsoids
must be triaxial and their excentricity must vary with radius at the
same time. Models (e.g. Madejsky \& M\"ollenhoff 1990) show that
even a moderate triaxiality can produce a considerable twist if
one looks at an elliptical galaxy under oblique view.

It is natural to expect the isophote twist due to projection
effects also in the case of galactic bars since they
are obviously triaxial and their excentricity shows a radial
variation (as seen in galaxies viewed face-on). In looking for
a correlation between central activity and isophote rotation,
one should separate the {\it intrinsic twists}, related to dynamics,
from mere {\it projection twists}.

The solution of such a task is outside the scope of this paper.
Nevertheless we would like to initiate the discussion on that topic
by several simple illustrative examples of projection effects on
artificially constructed single and double bars.

Fig. 1a shows
the ellipticity and PA profiles of a face-on viewed 2D bar,
whose isodensity
contours are perfect ellipses with axial ratio $a/b$ varying radially
from 1 to 3.
After projecting (with only moderate inclination, $I=30^o$) about the
line with $PA = PA_{proj}=0^o$ (i.e. coinciding with the bar minor axis),
the $PA$ along the bar becomes two-fold, with two plateaus separated by a
sharp $90^o$-transition at which the ellipticity falls locally
to zero (Fig.~1b). With the same $I$ but $PA_{proj}=30^o$, one obtains
a gradual twist of $\sim 50^o$ (Fig. 1c).

A 2D double barred system with
the inner component {\it perpendicular} to the outer one
is presented in Fig. 2a: the large-scale bar is the same as in
the above case;
the small one is 7 times shorter and its axial ratio $a/b$ varies
linearly from 1 to 2.
The projection with $I=60^o$ and $PA_{proj}=60^o$ is
shown in Fig. 2b: the local ellipticity maximum corresponding
to the secondary bar nearly disappears; the PA varies along both
primary and secondary bars.

Finally, a system of two {\it parallel} bars (with the same parameters as
above) is shown in Fig. 3a. The projection by $I=60^o$ about
the minor axis of both bars (i.e. $PA_{proj}=0^o$) results in an
illusion of two {\it perpendicular} bars (Fig. 3b).

The above examples clearly demonstrate that the projection
is a crucial factor for classifying twists and double bars.
To disentangle projection effects from intrinsic distortions, one
can try to deproject the observed images, making use of two advantages
spiral galaxies have with respect to ellipticals: a) they
are fairly two-dimensional except the bulge region and b) the inclination
$I$ and position angle $PA$ can be deduced from the shape of the outer disk
under the assumption that it is intrinsically circular. A two-dimensional
body with known $I$ and $PA$ can be deprojected without ambiguity:
if conditions a) and b) were strictly met, the problem would be solved.
Nevertheless many complications exist: the bulge is clearly tree-dimensional;
the primary bar may also be significantly thickened close to the center
due to the scattering on vertical resonances (e.g. Combes et al. 1990);
the secondary bar, when it exists, is confined to that bulge-bar
3D region; the outer disk has not necessarily the intrinsic circular
shape which can result in substantial errors in determining $I$ and $PA$.

%                                     One column figure
%_____________________________________2.
\begin{figure} [tbp]
%  \picplace{8cm}
%  \begin{center}
%\leavevmode
\epsfig{file=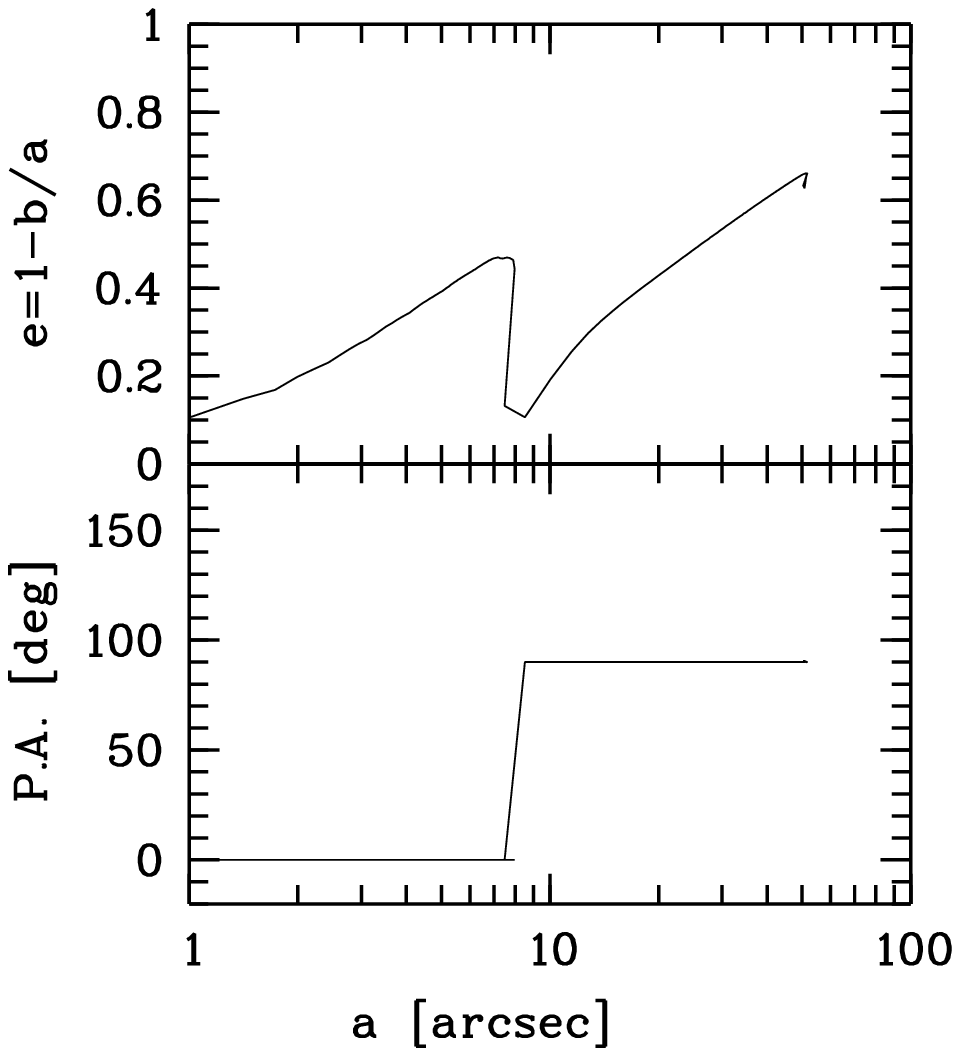, height=6.5cm, width=10cm, angle=-90}
\vskip-6.8cm
{\hskip+4cm
\epsfig{file=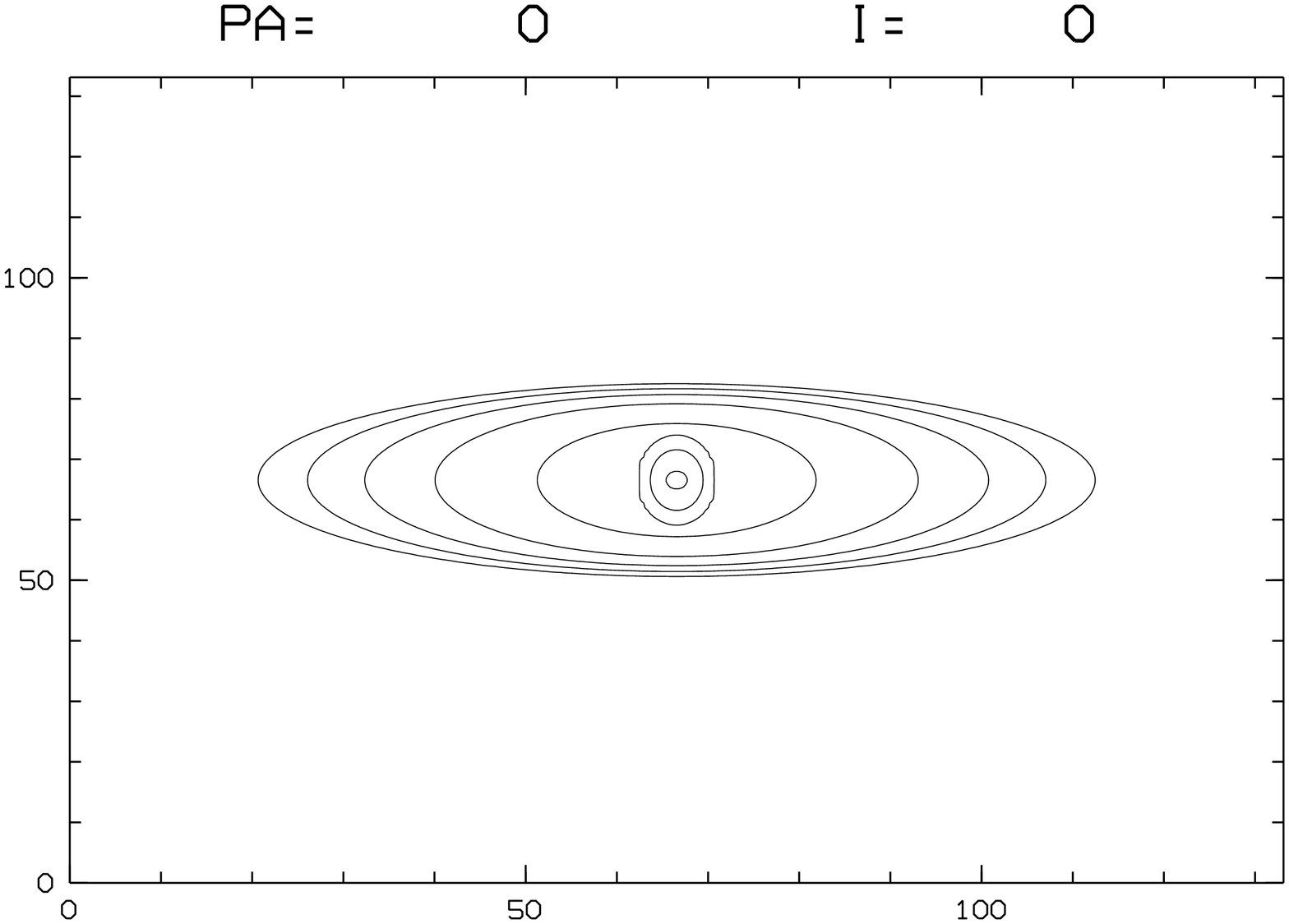, height=4.54cm, width=4.36cm, angle=-90}}
\vskip3mm
\epsfig{file=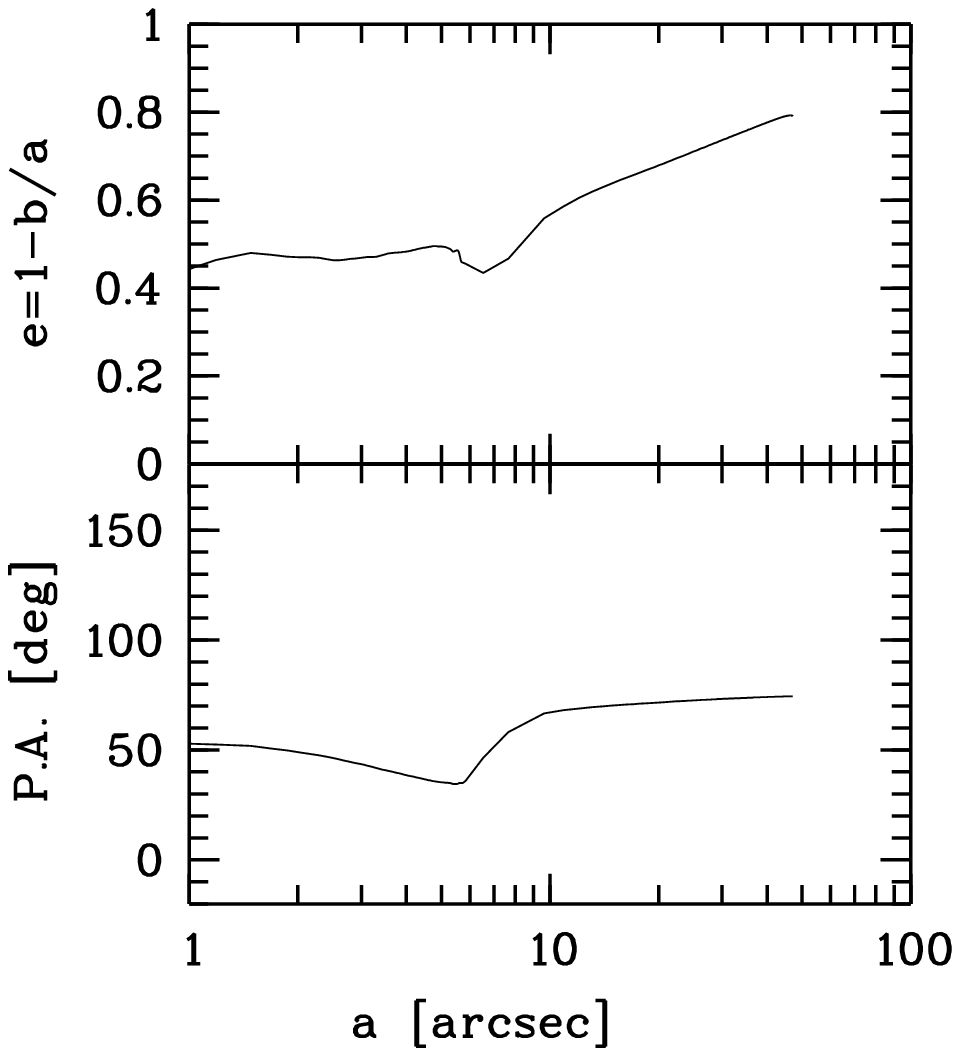, height=6.5cm, width=10cm, angle=-90}
\vskip-6.8cm
{\hskip+4cm
\epsfig{file=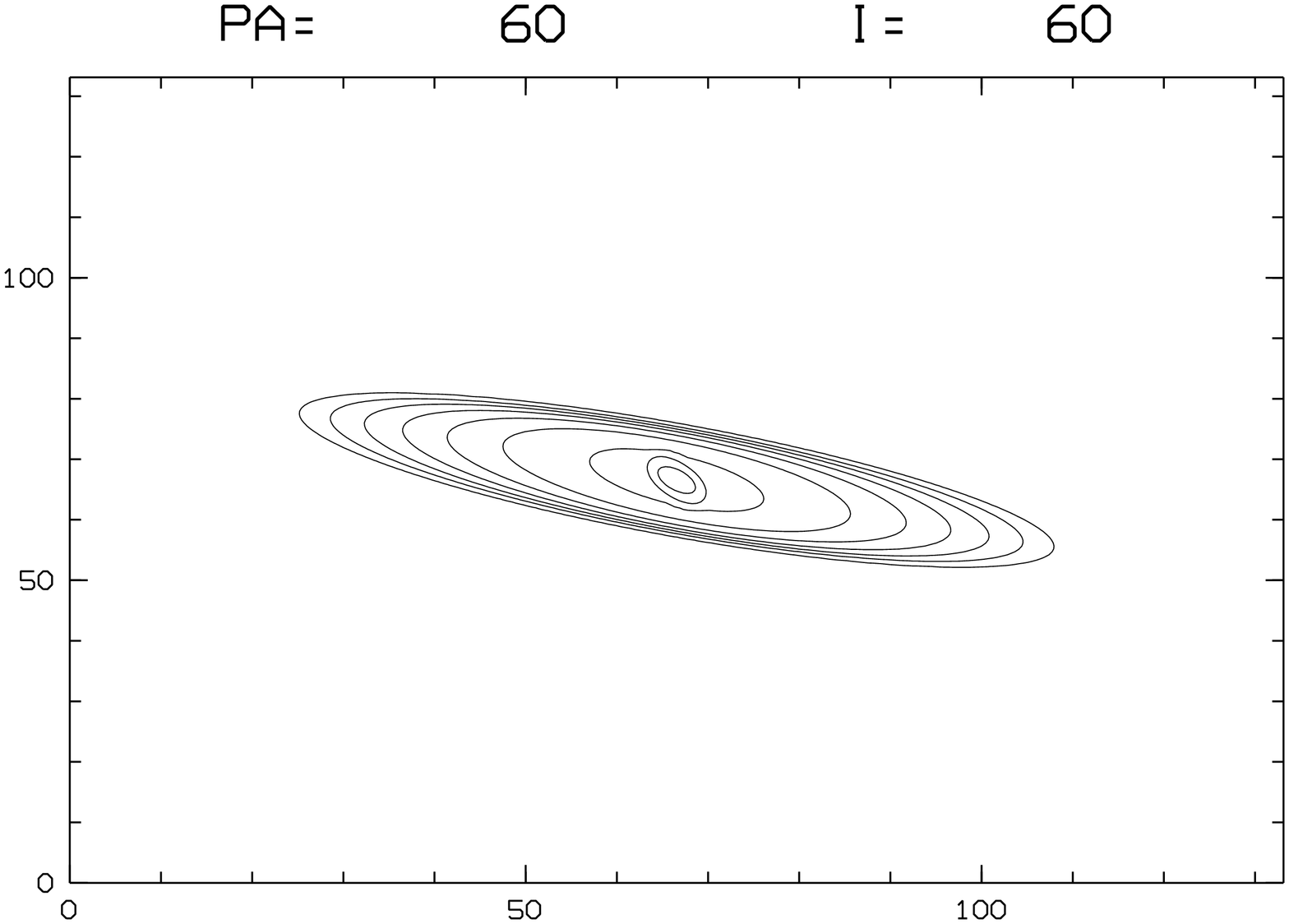, height=4.54cm, width=4.36cm, angle=-90}}
     \caption[]{
Ellipticities (1-b/a), position angles (PA) and contour plots for a
a 2D double bar -- bars perpendicular:
{\bf a)} Face-on view,
{\bf b)} Projection with $I=60^o$ and $PA_{proj}=60^o$}

%  \end{center}
\end{figure}
%--------------------------------------------------------------

%                                     One column figure
%_____________________________________3.
\begin{figure} [tbp]
%  \picplace{8cm}
%  \begin{center}
%\leavevmode
\epsfig{file=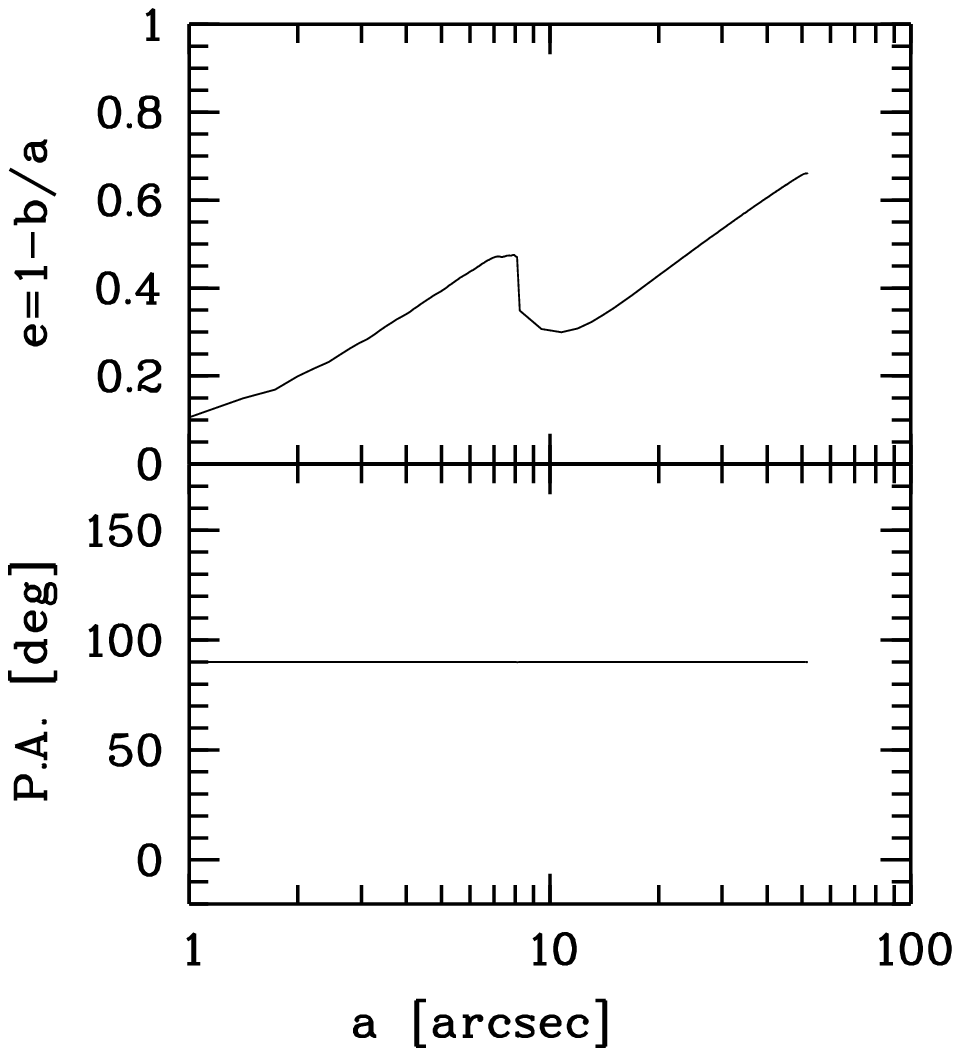, height=6.5cm, width=10cm, angle=-90}
\vskip-6.8cm
{\hskip+4cm
\epsfig{file=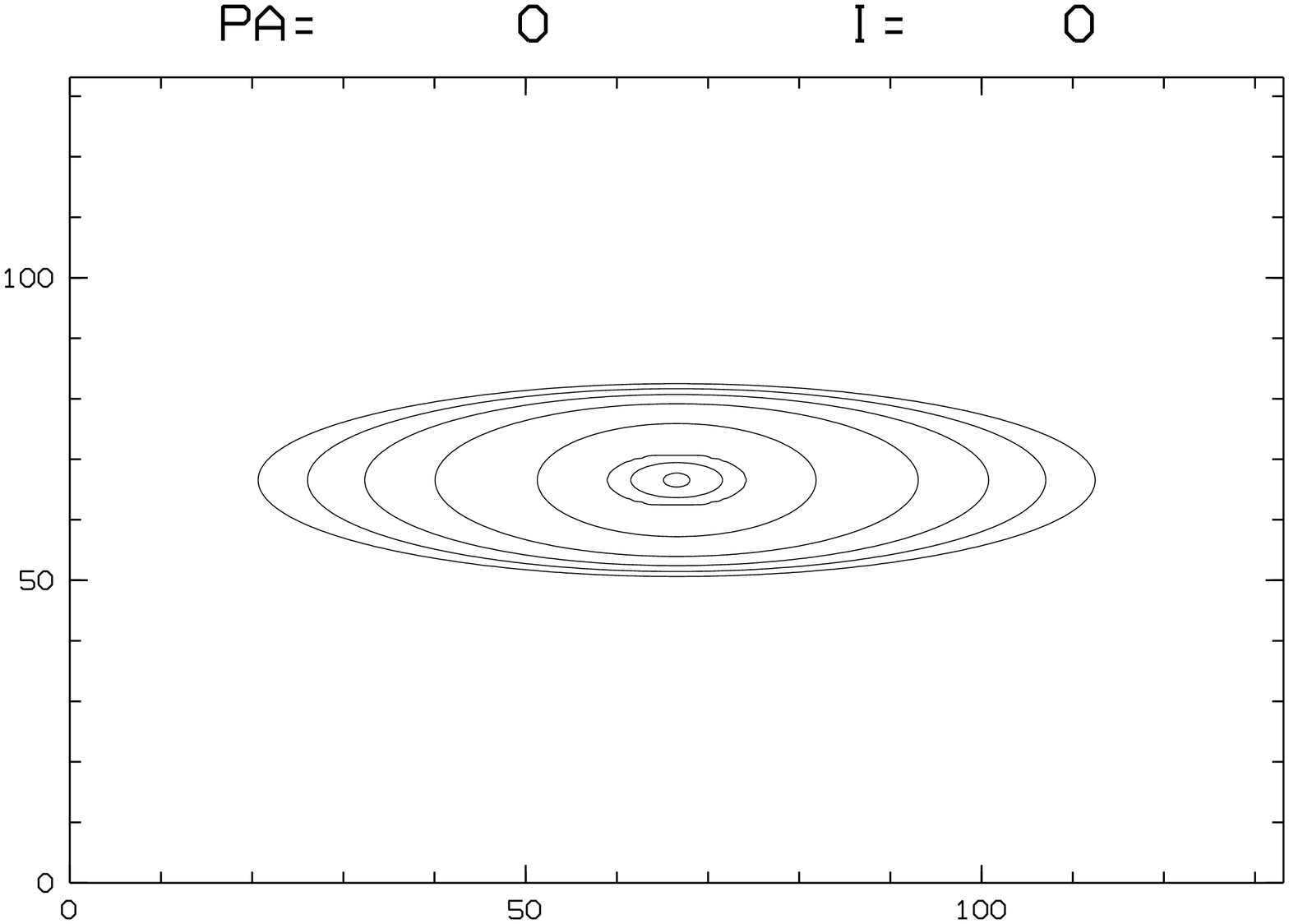, height=4.54cm, width=4.36cm, angle=-90}}
\vskip3mm
\epsfig{file=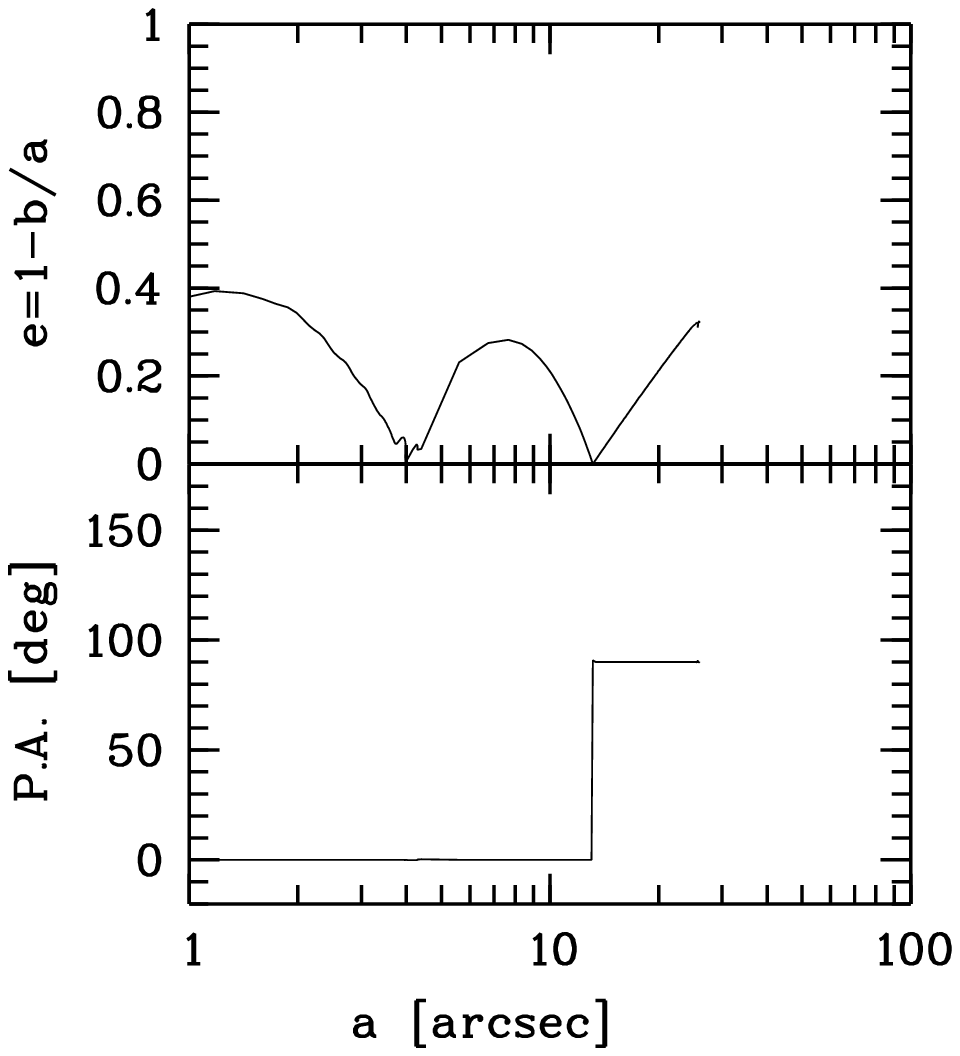, height=6.5cm, width=10cm, angle=-90}
\vskip-6.8cm
{\hskip+4cm
\epsfig{file=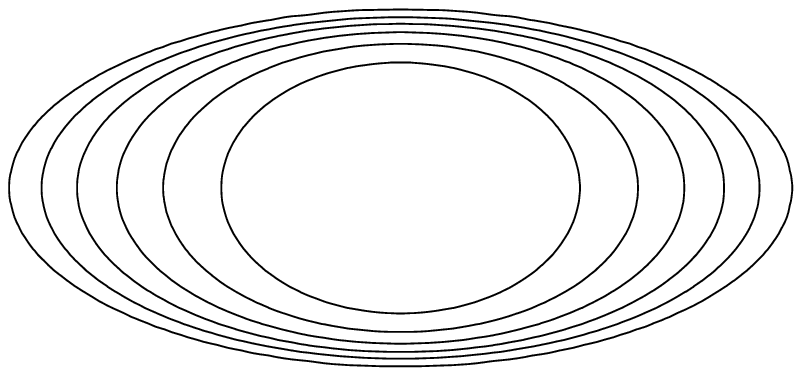, height=4.54cm, width=4.36cm, angle=-90}}
     \caption[]{
Ellipticities (1-b/a), position angles (PA) and contour plots for a
a 2D double bar -- bars parallel: {\bf a)} Face-on view,
               {\bf b)} Projection with $I=60^o$ and $PA_{proj}=0^o$,
                }
%     \caption[]{Isodensity contours of the axisymmetric + bar-like
%                distribution of matter in the plane $z=0$.}
%  \end{center}
\end{figure}
%--------------------------------------------------------------

In this paper, we have deprojected, under the assumption of
two-dimensionality, galaxies with inclination
lower than $75^o$, and we present, in the appendix, the deprojected
radial profiles together with the projected ones.
Since outer disks are usually located outside our images, we have used
disk inclination and position angles quoted in the Lyon-Meudon
Extragalactic Database (LEDA, Paturel et al. 1989).
%de Vaucouleurs et al. (1993) and, if not available there, from
%$I$ was derived from the axis ratio of the $25 mag/arcsec^2$ isophotal
%contour).

The deprojection can be done in two ways: either the image
is first deprojected and than a new ellipse fitting is carried out or
the ellipses fitted to the projected image are deprojected analytically.
The two approaches are not completely equivalent because of discreteness
of the detector array and because the isophotes are not perfect ellipses.
Our experiments have shown that the first method is less reliable:
after deprojecting the image, one has to interpolate to get intensity
at pixel positions which causes numerical errors resulting
in spurious variations of ellipticity and position angle in the
subsequent ellipse fitting.
We show this effect by dotted curves in Fig. 1a: they correspond to
the bar which is first projected with $I=30^o$ and $PA_{proj}=30^o$
(Fig. 1c) and than deprojected back to the face-on position. Both
ellipticity and position angle profiles significantly differ
from the correct ones (full lines) inside $a\sim 6''$, possibly giving
illusion of a small secondary bar. This numerical error is expected to
occur in regions with high density contrasts, e.g. close to the center
or at the edges of bars. Therefore we have preferred the second approach.

Whether the deprojected profiles are meaningful or not,
depends on how closely individual galaxies fulfill the conditions a) and b)
given above.
Clearly, in regions with non-negligible thickness, the error
resulting from the deprojection will grow with the galaxy inclination.
As can be seen from the profiles of observed galaxies, the
deprojection does not look reasonable in the bulge region if $I$
exceeds $\sim 45^o$.
Additional error is introduced by uncertainties in $I$ and $PA^{disk}$.

Being aware of big uncertainties in the deprojection procedure,
we do not rely on it to draw firm conclusions about the nature of a twist but
use it only as a secondary help:
if a double bar (or gradual twist) seen on the projected image
remains after deprojection, we consider the probability of its
existence to be strengthened; if it disappears, while $I<45^o$,
we take it to be a projection effect; on the other hand
if a double bar structure appears only after deprojection,
we do not classify it to be a double bar.
%\vfill\eject

\section{Individual galaxies}

%\command \subsection causes error !
%\subsection{Systems with two triaxial structures (double bars)}
\vskip1mm
{\noindent \it 4.1. Two triaxial structures (double bars)}
\vskip3mm

We include into this Section all galaxies which have, or at least
are suspected to have, more than one triaxial structure. We do not
strictly distinguish between double bars and bars with embedded
triaxial bulges since this difference is rather elusive.
Usually, we speak about double bars; only in cases when the
inner ellipticity maximum is low (after deprojection) and/or not
clearly separated from the outer one we use the
latter term. The classification of some galaxies as double-barred
is speculative, because either they are too inclined or the central
resolution is low.

\vskip3mm

%1. NEW DOUBLE BAR: NUCLEAR BAR ASSOCIATED WITH NUCLEAR SPIRAL
% VC96: Sey; Dev87
{\bf NGC 613} (SBbc/Seyfert, $1"\sim 86$ pc, $I=35.1^o$):\\
This Seyfert galaxy is known to possess, inside the large-scale bar,
an optical nuclear spiral. BC93 give semi-major
axes of the feature 7 x 6''. We identify short nuclear
spiral arms also
in the near-IR and suggest, on the basis of both
the ellipse fitting and the inspection of the grey-scale image, that
they have an associated nuclear bar ($e_{max}^s=0.55$ at $a=5.3"$).
If our interpretation is correct, then NGC 613 is a new example of a
double-barred system. The primary bar peaks ($e_{max}^p>0.72$) outside
our image.
%Interpretation inner spirals X sec. bar remains open.

% 2. NOTHING
{\bf NGC 1079} (RSAB0/a, $1"\sim 82$ pc, $I=57.1^o$):\\
The highest peak in ellipticity ($e_{max}=0.51$ at $a\sim 32"$) is
related to the large-scale bar, however the shape of the
profile can be deformed by the presence of spiral arms
at the end of the bar. There is a secondary
maximum ($e_{max}=0.32^s$ at $a=17''$) that we attribute
to the triaxiality of the bulge.

% 6. NEW DOUBLE BAR ?
{\bf NGC 1353} (SBb, $1"\sim 86$ pc, $I=70.2^o$):\\
The galaxy inclination is rather high and the interpretation therefore
uncertain: we suggest that the innermost ellipticity peak
is associated with a nuclear bar along which the PA varies as
a result of projection effects.
The large-scale bar is reflected by the PA plateau around
$a=14''$ and a corresponding small bump in the ellipticity.
The outermost ellipticity maximum is related to large-scale
spiral arms. The deprojection, although done,
is considered to be unreliable.

% VC96: Sey1;  Dev87;  Telesco93: Sey1/HII
{\bf NGC 1365} (SBb/Seyfert 1, $1"\sim 93$ pc, $I=58.1^o$):\\
%complicated structure, look to some paper
The galaxy has an IR-bright Seyfert nucleus and, as in the case of NGC 613,
a nuclear spiral, best seen as prominent dust lanes in optical
images (e.g. Teuben et al. 1986).
NGC 1365 is also classified as a starburst galaxy (e.g. Telesco
et al. 1993), with the star-forming activity concentrated in
circumnuclear ``hot-spots''.

The morphology of the nuclear region is complex and patchy also
in our H image,
indicating that the emission of old red stars is probably strongly
contaminated by the light of new red giants and supergiants formed
in the starburst. The nuclear spiral is well recognized and we identify a
nuclear bar embedded in it:
the peak in the ellipticity ($e_{max}^s=0.46$ at $a=8.3$) is
related to that bar while the adjacent minimum in the PA
is related to the nuclear spiral.
The nuclear bar is roughly parallel (NE-SW)
to the elliptical distribution of circumnuclear molecular
gas mapped by Sandqvist et al.(1995).
Neither the spiral nor the bar are smooth, unlike in NGC 613.

% 8. TWIST EXPLAINED BY PROJECTION EFFECTS (TRIAXIALITY)
% W95
{\bf NGC 1398} (R'SBab, $1"\sim 80$ pc, $I=48.0^o$):\\
The disk is dominated by a well defined large-scale bar
reaching  $e_{max}$ (0.37) at $a=36"$ after which it passes into
an outer
ring. The inner isophotes ($a<20"$, $PA\sim 80-90^o$)
are not aligned with the primary
bar ($PA^p \sim 12^o$) and are slightly twisted
(by $\sim 13^o$ between 5 and 12"). A small bump in profiles near a=14"
was found also by W95 but they were reluctant to interpret it.
We consider it is related to the triaxiality of the bulge; its
signature is seen also in deprojected profiles ($PA_{disk}=100^o$).
Note that the inner isophote twist almost disappears after deprojection:
the PA is constant (within $7^o$) along the
whole 60"-profile.

% 10. KNOWN DOUBLE BAR
% Buta 86, W95
{\bf NGC 1433} (R'SBab, $1"\sim 53$ pc, $I=24.5^o$):\\
The double barred structure of the galaxy was
already reported by Buta (1986) and W95 in BVRI bands.
We find, in JHK bands, the length
$l_{max}^s$ of the secondary bar to be 5.6, 5.4 and 6.2''.
For comparison with W95 we measured also $l_{min}^s = 12.9, 13.0$
and 12.8'' (W95 give 11.5" in filter I).
The primary bar exceeds in
length our frame, so that its ellipticity is still increasing
at the last point of the profile. The PAs of bars
are 32 and $95^o$ (with the nuclear bar leading),
in good agreement with W95 (30 and 94$^o$).
%No deprojection

%12 SUSPECTED TWIST
{\bf NGC 1512} (SBa, $1"\sim 43$ pc, $I=62.9^o$):\\
The image is dominated by a large-scale bar showing
isophotal twist ($\Delta PA \sim 30^o$) exterior to $a\sim 7"$
(noticed by E96 on the blue plate). A low ellipticity maximum
and the associated change of the PA at $a\sim 6''$ can be indicative
of a distinct component, probably a triaxial bulge. This interpretation
is however uncertain because of high inclination (note a partial similarity
with Fig. 2 showing a projection of a double barred system).

% VC96: HII; Telesco93
{\bf NGC 1808} (RSABa/HII, $1"\sim 50$ pc, $I=60.6^o$):\\
NGC 1808 is a nearby starburst galaxy (e.g. Telesco 1993) with
many star forming knots inside 1 kpc. It is also suspected to
possess a hidden Seyfert nucleus (e.g.V\'eron-Cetty \& V\'eron 1985).

On the H image the nuclear region looks smooth, unlike in optical.
This is consistent with Tacconi-Garman et al. (1996) who claim
that the near-IR emission from the nucleus of this galaxy is dominated
by old stars with only a small ($<10\%$) contribution from
young red giants and supergiants born in the starburst.
We interpret the peak in the excentricity and the associated PA plateau
at $a\sim 3''$ as a manifestation of a nuclear bar (its presence
was recently deduced also by Kotilainen et al. (1996) on the basis of
JHK contour plots).
We do not attempt to give parameters of the large-scale bar since
the galaxy has a peculiar morphology (possibly due to the interaction
with NGC 1792) and our image is spatially rather small. Phillips
(1993) classifies the galaxy as barred on the basis of the distribution
of HII regions and Saikia et al. (1990) report an HI bar 22 kpc long.
Therefore we classify NGC 1808 as being double-barred.

%20 NEW DOUBLE BAR
{\bf NGC 2217} (RSB0+, $1"\sim 91$ pc, $I=28.6^o$):\\
We identify this low-inclination galaxy as a new double-barred system:
the two
maxima in ellipticity ($e_{max}^s=0.19$ at $a=7.8"$ and $e_{max}^p=0.48$
at $a=37"$ occur on the approximate plateaus of the PA ($138^o$
and $112^o$ for the secondary and primary bar, respectively).
% depro not done, PA unknown

%28 NEW DOUBLE BAR ?
%Dev87; Elm96 (twist on B images)
{\bf NGC 2935} (R'SABb, $1"\sim 136$ pc, $I=43.4^o$):\\
The outer ellipticity peak corresponds to the
large-scale bar ($e_{max}=0.52$ at $a=25"$). We interpret the secondary
maximum, still perceptible after deprojection,
as being related to another triaxial component, probably
the bulge (the shape of the profile in the transition region
between the two peaks can be
partly deformed by the presence of a badly masked star near
the major axis of the inner component at $a\sim 11''$, however
the inspection of the grey-scale image confirms that the
inner misaligned structure really exists).
The smooth twist of the barred isophotes (already noticed
by E96 on blue plates) is explainable
by projection effects. After deprojecting, the PA is two-fold with
two approximate plateaus.

%33 NEW DOUBLE BAR
{\bf NGC 3368} (SABab, $1"\sim 59$ pc, $I=51.0^o$):\\
One can see three ellipticity maxima ($e_{max}=0.29$, 0.30 and
0.43) accompanied by three plateaus in the PA profile. The third
maximum is controversial since it
is close to the locus where spiral arms start and our frame ends at the
same time.
The illusion of the triple-barred system disappears after deprojection
($PA_{disk}=5.0^o$): it seems
plausible that the galaxy is double-barred with the two bars roughly
aligned. However, as $I$ is rather high for the deprojection to be
unambiguous, we do not exclude that three triaxial components coexist.

%35 TWIST or SMALL SECONDARY BAR
%VC96: Sey 2;
{\bf NGC 3393} (R'SBa/Sey 2, $1"\sim 230$ pc, $I=23.7^o$):\\
Inside the large-scale bar ($e_{max}=0.46$ at $a=13.3$),
there is an isophote twist of $\sim 11^o$ between $a=3"$ and $4.4"$
We speculate that it could result from the existence
of a small secondary bar, since there is a local maximum
of ellipticity near $a=2"$, while the PA has a plateau there.
The proximity of that region to the center makes
the double-bar classification uncertain;
higher resolution is needed to confirm
the discovery.

%68 NEW DOUBLE  BAR
{\bf NGC 4984} (RSAB0+, $1"\sim 73$ pc, $I=45.9^o$):\\
The ellipticity peak ($e_{max}=0.23$)
around $a=4"$ is associated with a short
plateau in the PA ($64^o$ between 2.9 and 4.1"). Since the
peak is even better seen after deprojection ($PA_{disk}=15^o$),
we believe that it indicates the presence of a short bar.
The outer ellipticity maximum ($e_{max}=0.30$ at $a=30''$) is
related to the primary bar.
Both bars are intrinsically nearly parallel provided the deprojection
is correct (cf. Fig.~3).
% the nature of the second peak is unclear. it is probably
% related to spiral arms but I dont see any !

%69 STRONG TWIST
{\bf NGC 5101} (RSB0/a, $1"\sim 110$ pc, $I=33.6^o$):\\
The isophotes inside the large-scale bar
($e_{max}=0.56$ at $a=50"$)
are strongly twisted ($\Delta PA \sim 106^o$). Such a twist is
not explainable by mere projection effects in view of rather
low galaxy inclination. Either there is an intrinsic gradual twist
or a nuclear bar as would suggest the ellipticity
maximum and the PA plateau at $a<3''$: however the proximity
of this feature to the center
as well as its low ellipticity ($e_{max}=0.06$ at $a=2''$) make
this hypothesis very speculative.
% depro not uvailable

{\bf NGC 5566} (SBab, $1"\sim 100$ pc, $I=79.5$):\\
In spite of the high inclination of the galaxy, we interpret the profiles
in terms of a double barred structure with the higher maximum in ellipticity
($e_{max}^s=0.56$ at $a\sim 6''$) corresponding to the nuclear bar.
The large-scale bar has, due to projection, only a low
ellipticity peak ($e_{max}^p=0.24$ at $a\sim24''$) and a short PA dip.
We think that the deprojected profiles are not meaningful -- because
$I\sim 80^o$ -- and show them only as illustration of how weak the
deprojection procedure is for such high inclinations (e.g. note
the constancy of the deprojected ellipticity
at a high value near the center).

It is interesting to compare this galaxy with NGC 3166 since
both are highly inclined and their morphology is apparently
similar on undeprojected images: at first sight, both seem to
show a double barred structure with
the components roughly perpendicular. Nevertheless, in the light of
Section 3 and Figs. 1 and 3, we classify NGC 3166 as having only a
large-scale bar while NGC 5566 is claimed to have two bars. To
grasp the difference, note that in the (undeprojected)
ellipticity profile of NGC 3166, there is no local maximum interior
to that of the large bar; rather, the ellipticity monotonically
climbs towards the center. This is characteristic of
projection effects: the behaviour is qualitatively similar to Fig. 1b
in which the bar is projected along its minor axis as is the
case of NGC 3166 (in comparison, keep in mind that the profile of
NGC 3166 reflects also the disk -- which adds the change of ellipticity
and PA rightwards from the bar ellipticity maximum  --
and that the inclinations are different --
which changes the relative height of the peak).
On the other hand, NGC 5566 shows two ellipticity peaks (excluding
the outer raising slope related to the disk); ellipticity decreases
towards the center. That is why we interpret the structure as double
barred: no projection of a single bar with reasonable ellipticity profile
(monotonically increasing as suggested by face-on single barred
galaxies) can generate a secondary ellipticity peak.

% 9. KNOWN DOUBLE BAR
% W95
{\bf NGC 6782} (RSABa, $1"\sim 230$ pc, $I=44.4^o$):\\
A double bar structure, discussed already by BC93 and W95,
exists, with the primary ($e_{max}^p=0.51$, $l_{max}^p=26"$)
and secondary ($e_{max}^s=0.36$, $l_{max}^s=3.4"$) bars
oriented at $PA=178^o$ and $150^o$, respectively.
The deprojection ($PA_{disk}=45^o$) does not alter this picture:
two peaks in ellipticity, along which the PA is roughly constant,
still exist.

%75 NEW DOUBLE BAR
{\bf ESO 437- 67} (R'SBab, $1"\sim 190$ pc, $I=29.9^o$):\\
This galaxy could be double-barred since
two peaks in ellipticity ($e_{max}=0.23$ at 2.5" and 0.62 at 31.8")
are associated with approximate plateaus in the PA (139 and $119^o$).
Nevertheless, this classification is speculative because the
inner feature is rather close to the center: higher resolution
is desirable to confirm the hypothesis.

%\vfill\eject

%------------------------------------------------------------------
\begin{table}
\caption{Projected parameters of double bars} \label{tbl-1}
\vskip-5mm
\begin{center}\scriptsize
\begin{tabular}{llrrrrrr}
Galaxy &$e_{max}^s$ &$l_{max}^s$ &$PA^s$ &$e_{max}^p$ &$l_{max}^p$ &$PA^p$ \\
       &            & ('')       &($^o$) &            &('')        &($^o$) \\
%\multicolumn{1}{c}{$P$\tablenotemark{a}} & $P R_{maj}$ & $P R_{min}$ &
%\multicolumn{1}{c}{$\Theta$\tablenotemark{b}} \\
%\tableline
     &    \\
N 613     & 0.55 & 5.3 & 122 & $>0.72$ & $>59$ & 127 \\
N 1079    & 0.32 & 17  & 96  & 0.51    & 32    & 122 \\
N 1365    & 0.46 & 8.2 &  46 & ?       & ?     &   \\
N 1353    & 0.40 & 3.6 & 131 & 0.31    & 14    & 184 \\
N 1398    & 0.14 & 14  &  83 & 0.37    & 36    & 12  \\
N 1433(H) & 0.37 & 5.4 & 32  & ?       & ?     & ?  \\
N 1433(J) & 0.35 & 5.6 & 32  & ?       & ?     & ?  \\
N 1433(K) & 0.36 & 6.2 & 32  & ?       & ?     & ?  \\
N 1512    & 0.13 & 5.7 & 74  & 0.61?   & 68?   & 45  \\
N 1808    & 0.53 & 3.3 & 158 & ?       & ?     & ?   \\
N 2217    & 0.19 & 7.8 & 138 & 0.47    & 37    & 112 \\
N 2935    & 0.32 & 10  & 153 & 0.52    & 25    & 134 \\
N 3368    & 0.29 & 4.2 & 130 & 0.30    & 24    & 156 \\
N 3393    & 0.20 & 2.0 & 147 & 0.46    & 13    & 159 \\
N 4984    & 0.23 & 4.0 & 64  & 0.30    & 30    & 95  \\
N 5101    & 0.06 & 2.0 & 46  & 0.56    & 50    & 121 \\
N 5566    & 0.56 & 5.9 & 37  & 0.24    & 24    & 156 \\
N 6782    & 0.36 & 3.4 & 150 & 0.51    & 26    & 178 \\
E 437-67  & 0.23 & 2.5 & 138 & 0.62    & 32    & 119 \\
\end{tabular}
\end{center}
%(a) from optical, Martin (1995)
\end{table}
%------------------------------------------------------------------

%\command \subsection causes error !
%\subsection{SA galaxies with a nuclear bar}
\vskip3mm
{\noindent \it 4.2. SA galaxies with a nuclear bar}
\vskip3mm

%17 TWIST DISAPPEARS AFTER DEPRO
{\bf NGC 1792}  (SAbc, $1"\sim 64$ pc, $I=62.0^o$):\\
The ellipticity maximum inside 10" is related to
a short nuclear bar ($e_{max}=0.45$ at 5") along which no clear twist
occurs. Further ellipticity maxima are connected to
spiral arms.
% galaxy class. as SBbc but the ell. has no clear max. after
% deprojecting, so isn't ther just a triax. bulge rather than a bar ?
% Also, on proj. image, the bar is rather short (proj. maximum
%at 320 pc for dist=13 Mpc

{\bf NGC 4438} (SA0/a(pec)/Sey 3, $I=78^o$):\\
This galaxy (Arp 120), possessing a Seyfert nucleus,
is interacting with a close companion
NGC 4435 in the Virgo cluster
Its nuclear regions might be perturbed by infalling gas
(e.g. Combes et al 1988, Kenney et al 1995).
Its inner maximum in ellipticity could reflect either a tidally
deformed bulge or a nuclear bar.

% 7. TWO CONTROVERSIAL INNER OVALS
% W95, VC96: Sey2, Arp(interacting galaxy)
{\bf NGC 5427} (SAc/Seyfert 2, $1"\sim 170$ pc, $I=38.9^o$):\\
The ellipticity profile shows four clear maxima, the highest
of which (at $a\sim 21"$) corresponds to the two-arm spiral structure.
The two innermost ones ($a\sim 5"$ and $8"$) do not correspond
to well defined plateaus in PA, however the changes in ellipticity
and PA occur at roughly the same places.
We agree with W95 that these features are oval structures
whose nature remains controversial, which is perhaps connected to the
interacting nature of NGC 5427.
Nevertheless, we note that after
deprojection, the PA is defined much better in the inner region:
it is constant up to 6", i.e. within the first ellipticity
peak which could indicate a nuclear
bar in this region. This interpretation is tempting also
because of the Seyfert nucleus.
%\vfill\eject

%\command \subsection causes error !
%\subsection{Twisted bar isophotes}
\vskip3mm
{\noindent \it 4.3. Twisted bar isophotes}
\vskip3mm

In this section, galaxies with gradual twist along their
large-scale bars are described. The selection is done on the basis
of undeprojected images: some of such twists are likely not to
be intrinsic.
\vskip3mm

% 3. DOUBLE BARRED AFTER DEPROJECTION ?
% Dev87
{\bf NGC 1187} (SBc, $1"\sim 82$ pc, $I=46.7^o$):\\
The ellipticity grows to its maximum
($e_{max}=0.60$) at $a=32"$, reflecting the large-scale bar from
the tips of which two-arm spiral structure emanates.
The PA is roughly constant inside the bar except the innermost
region ($a=2-7"$) where a $15^o$-twist occurs.
After the deprojection ($PA_{disk}=130^o$), a possibility
that the system is double-barred emerges: the PA is essentially
two-fold and there is a new maximum in ellipticity
at about 3".
In case that the secondary bar
really existed, it would be nearly perpendicular to the primary
one ($\Delta PA=87^o$).
However, we believe that such an appearance
is likely to be an artefact of a wrong deprojection (cf. Fig. 1)
and, in accordance with rules established in Sect. 3,
we classify the galaxy only as having a twist.

% 5. SMALL TWIST INSIDE THE BAR
% W95
{\bf NGC 1302} (RSB0/a, $1"\sim 96$ pc, $I=27.4^o$):\\
Between $a=3-7"$ the PA is roughly constant
($180-190^o$) but then it turns by $20^o$
along the bar whose $e_{max}$ (0.35) occurs at $a=29"$. This
twist was already noted by W95 who attributed it to
the presence of dust lanes since its amplitude depends
on the band (BVRI). It is important that we found it
also in band H where the dust extinction is much less
important.
% compare twist amplitude with W95.
% No deprojection

% 9. ELLIPTICAL
%{\bf NGC 1404} (E, $1"\sim 110$ pc, $I=49.3^o$):\\
% Elliptical galaxy. No isophotal twist detected.

%11 STRANGE - WRONG ?
%{\bf NGC 1510} (S0, $1"\sim 48$ pc, $I=67.9^o$):\\
%No evidence for twist found, neither after deprojection.
% Strange: - clasified as S0 but I get $e=0.6$ after deprojection
%          - appears very small while it is close (10 kpc): noisy
% Eliminate ?

%19 TWIST
% Dev87;
% Rings (is there 4/1 ring ?)
{\bf NGC 1832} (SBbc, $1"\sim 115$ pc, $I=46.3^o$):\\
The excentricity of the large-scale bar grows
continuously to $e_{max}=0.58$ at $a=17"$ where a regular two-armed
spiral structure starts. There is a twist of amplitude $14^o$
inside the bar. The twist is even more pronounced
($\Delta PA = 53^o$) after deprojection
($PA_{disk}=10^o$).

%24 TWIST
{\bf NGC 2442}  (SABbc, $1"\sim 74$ pc, $I=27.8^o$):\\
The inner isophotes are twisted inside the large-scale
bar: $\Delta PA \sim 29^o$ between $a=3"$ and 10".
% depro not done, PA unknown

%25 TWIST ?
{\bf NGC 2525} (SBc, $1"\sim 93$ pc, $I=48.2^o$):\\
The large-scale bar, with maximum ellipticity of 0.69
(at $a=22"$), has twisted isophotes: $\Delta PA = 13^o$. The
deprojection ($PA_{disk} = 75^o$)
is suspicious: the resulting big twist is probably the consequence
of unexact deprojection parameters and noisy image.

%29
{\bf NGC~2997} (SABc, $1"\sim 57$ pc, $I=44.5^o$):\\
The twist of amplitude $25^o$ between $a=4''$ and 7.3"
can be due to projection effects alone.
% I see no clear evidence for a bar after deprojection ==>
%   is there just a triax. bulge

%30
% VC96: Sey1
%{\bf NGC 3080} (Sa/Seyfert 1, $1"\sim 670$ pc, $I=21.4^o$):\\
%very far (140 Mpc) ==> low resolution (1 pix = 350 pc)
%          to follow twisting

%31 NOTHING AFTER DEPROJECTION (before: double bar)
% Dev87
{\bf NGC 3166} (SAB0/a, $1"\sim 84$ pc, $I=73^o$):\\
Looking at inner contours, one could easily
get impression that the galaxy is double-barred, with the bars roughly
perpendicular one to another. However, this appearance
is probably entirely due to projection effects (note that
no ellipticity peak corresponds to the first PA plateau; cf. Fig. 1
and comments on NGC 5566). After deprojection
($PA_{disk}=87^o$), no clear isophote twist is found.
% Strange: gal. class. as S0-a but it is very elongated
% after deprojection ! Is it not an elliptical gal ? Or, are the proj.
% parametres wrong ?

%37 STRONG TWIST
{\bf NGC 3637} (RSB0/a, $1"\sim 115$ pc, $I=30.3^o$):\\
Inside the large-scale bar ($e_{max}=0.38$ at $a=14.6"$; $PA=37^o$),
a strong twist (amplitude $\sim 71^o$ in the region $a=3"-14.6"$) is
measured. The resolution in the inner part is not sufficient
to say whether a secondary bar exists.
% depro unavailable; I=30;

%38 SMALL TWIST
{\bf NGC 3673} (SBb, $1"\sim 115$ pc, $I=51.6^o$):\\
There is a small twist ($17^o$ between $a=3"$ and 42") inside
the large-scale bar ($e_{max} = 0.73$ at a=42")

%40 TWIST
% Dev87;
{\bf NGC 3887} (SBbc, $1"\sim 71$ pc, $I=39.5^o$):\\
Inner isophotes (between $a=3"$ and 25") inside
the large-scale bar ($e_{max}=0.68$ at $a=33"$; $PA=181^o$)
are twisted by $22^o$.

%42B
%NGC 4106B:

%48
% Dev87
%{\bf NGC 4273} (SBc, $1"\sim 150$ pc, $I=52.9^o$):\\
% complicated structure & low resol. (distance=32 Mpc) - look to ref.
% multiple peaks, but not clear whether related to bars
% depro suggests 2 inner spiral arms emanating from a small bar
% Att: one half is much brighter, be careful !

%52 TWIST ? (DISAPPEARS AFTER DEPRO)
{\bf NGC 4454} (RSB0/a, $1"\sim 140$ pc, $I=29.5^o$):\\
There is a strong twist ($\Delta PA=52^o$ between $a=3"$
and $12"$) inside the large-scale bar ($e_{max}=0.53$ at $a=31"$;
$PA=22^o$). The galaxy is only moderately inclined ($I=30^o$)
and the twist almost disappears (reduces to $\Delta PA = 10^o$)
after deprojection.

%58 TWIST; NEW BAR ?
{\bf NGC 4612} (SB0+, $1"\sim 120$ pc, $I=47.1^o$):\\
The isophotes of this lenticular galaxy severely
twist ($\Delta PA=48^o$) between $a=2"$ and $a=15.7"$ where
the ellipticity has a maximum ($e_{max}=0.22$). After deprojection
($PA_{disk}=145^o$), the ellipticity has
a peak ($e_{max}=0.4$) along which the PA is roughly constant
while a small twist
($\Delta PA=14^o$) still exists within the innermost 10".
We interpret the profiles as resulting from the presence
of a weak bar (we do not feel obvious to classify
the galaxy as strongly barred as it is in RC3).
% Att: galaxy has an outer ring ! Is not the max. in ell.
% conn to it rather than to a bar ? (the max is quite far:
% 16" on proj. and 25" on depro: chech the ring diameter in ref.)

%59 TWIST; PROBLEM
{\bf NGC 4665} (SB0/a, $1"\sim 51$ pc, $I=2.6^o$):\\
There is a small gradual twist along the large-scale bar
($11^o$ between a=3" and 42" where).

%72 STRONG TWIST
%VC96: Sey2
{\bf NGC 5643} (SABc/Seyfert 2, $1"\sim 66$ pc, $I=28.8^o$):\\
Despite the low disk inclination,
there is a strong twist ($\Delta PA=44^o$ between
$a=3"$ and 30") inside the large-scale bar. This is interesting,
because NGC 5643 belongs to late-type (Sc) galaxies -- which are
expected not to show a twist (Elmegreen et al. 1996) -- and, at the
same time, has a Seyfert nucleus.
% depro not available, i=29

%73 STRONG TWIST
{\bf NGC 5701} (RSB0/a, $1"\sim 100$ pc, $I=15.2^o$):\\
There is a strong twist ($\Delta PA=39^o$ between
$a=3"$ and 38") inside the large-scale bar ($e_{max}=0.44$
 at $a=38"$,
$PA=0^o$), probably not explainable by projection effects since
the galaxy inclination is only 15$^o$.
% depro unavailable, i=15.2

%\command \subsection causes error !
%\subsection{No nuclear twist or complex morphology}
\vskip3mm
{\noindent \it 4.4. No nuclear twist or complex morphology}
\vskip3mm

% 4. % NORMAL BAR
{\bf NGC 1255} (SABbc, $1"\sim 96$ pc, $I=51.4^o$):\\
No evidence for twist is found; the first, rather flat peak
in ellipticity ($e_{max}=0.53$ at $a=4.4"$) corresponds to the bar
($PA=116^o$) , while
the second, more pronounced (at $a\sim 30"$) is related to a double-armed
spiral structure. The profiles are
qualitatively similar after deprojection ($PA_{disk}=117^o$).

%14 NO TWIST
{\bf NGC 1640} (SBb, $1"\sim 92$ pc, $I=29.1^o$):\\
The PA is constant ($\sim 45^o$) along the large-scale
bar whose ellipticity maximum (0.62) is reached at $a=31"$.
% cf. 1187 where a double bar appears after deprojection

%15 NOTHING
{\bf NGC 1744} (SBd, $1"\sim 36$ pc, $I=64.4^o$):\\
The galaxy is inclined at $I\sim 65^o$ and, moreover, our image is
noisy: the ellipse
fitting is not reliable. No clear evidence for twist is found.

%16 NOTHING
% Dev87;
{\bf NGC 1784} (SBc, $1"\sim 140$ pc, $I=53.2^o$):\\
 The ellipticity profile shows one maximum ($e_{max}=0.68$ at
$a=26"$) near the end of the large-scale bar. The PA is well constant
(between 91 and $95^o$) along the bar.
% outer ring seen on deprojected image

%27 NOTHING
% VC96: Sey 3
{\bf NGC 2911} (SA0/Seyfert 3, $1"\sim 200$ pc, $I=47.4^o$):\\
The PA, as well as the ellipticity, are rather flat inside
$a=12"$. The small features further out are most likely
due to badly masked stars.

%32
{\bf NGC 3346} (SBcd, $1"\sim 82$ pc, $I=20.1^o$):\\
No twist is found inside the primary bar ($e_{max}=0.69$
at $a=13"$).
% depro parameters unavailable

%41 NOTHING
{\bf NGC 4050} (SBab, $1"\sim 115$ pc, $I=46.8^o$):\\
The PA is roughly constant ($77-87^o$)
along the large-scale bar.
% twist after deprojection, probably artificial

%42 NOTHING
{\bf NGC 4106} (SB0+, $1"\sim 130$ pc, $I=51.0^o$):\\
The PA is remarkably constant ($90-93^o$)
within $a=11.5"$. The ellipticity has a maximum (0.37) inside
that PA plateau (at $a=5.3"$). The PA grows later on while the
ellipticity decreases.
% discuss deprojection - interesting hole in PA
% discuss in connection to NGC4106B - are related ?

%45 NO BAR, NO TWIST
%Dev 87;
{\bf NGC 4212} (SAc, $I=51.5^o$):\\
The ellipticity peak near $a=29"$ is related to spiral
arms, not to a bar structure. The inner isophotes, inside 10",
show no clear twist: an oblate bulge is probably sufficient to explain
the profiles.
% depro: interpretation unclear: oval/triax. bulge/wrong depro

%47
{\bf NGC 4267} (SB0-, $1"\sim 71$ pc, $I=22.5^o$):\\
The PA is constant along the large-scale bar
($e_{max}=0.22$ at $a=18.4"$, $PA=32^o$) of this low-inclination
galaxy.
% depro unavailable
% E-SO

%49 NO TWIST, HIGH INCLINATION
{\bf NGC 4424} (SBa, $I=66.8^o$):\\
The galaxy is inclined at $I=67^o$.
The twist inside the bar ($e_{max}=0.78$ reached at $a=8"$;
$PA=110^o$) does not exceed $7^o$.
% small companion at N ? what is it

%54 NEW WEAK BAR ??
% Dev87; VC96: Sey2
{\bf NGC 4501} (SAb/Seyfert 2, $1"\sim 150$ pc, $I=60.2^o$):\\
The bulge region is characterized by growing ellipticity.
However, the constancy of the PA ($140-144^o$ interior
to $a=18"$) -- as well as the fact that its value corresponds to that of
the outer disk -- is compatible with a spheroidal shape of the bulge:
there is no evidence for a bar or other triaxiality.
The change of excentricity and PA behind $a=18"$ reflects
the presence of spiral arms.
% inner disk excentric after depro ==> oval/bar/triax. bulge ?

%56
{\bf NGC 4519} (SBd, $1"\sim 80$ pc, $I=36.5^o$):\\
No isophotal twist is found inside the bar region
of this late-type galaxy.
% after depro: no twist neither
% before depro: peak in exc. at a=3" (too close to center ?)

%61 TWIST (TRIAXIAL) ?
{\bf NGC 4689} (SAbc, $1"\sim 110$ pc, $I=33.0^o$):\\
Multiple features in the ellipticity and PA profiles result
from flocculent spiral structure. No clear twist occurs
in the nuclear region.
% depro not available (I=33)

%63 INCLINED, UNRELIABLE
% Dev87
{\bf NGC 4731} (SBcd, $1"\sim 94$ pc, $I=69.3^o$):\\
The galaxy is rather inclined and the ellipse
fitting is unreliable. No clear twist is found.
% depro gives very strange output so don't speak about it

%64
% Dev87
{\bf NGC 4781} (SBd, $1"\sim 77$ pc, $I=67.9^o$):\\
No clear twist is found inside the large-scale bar
($e_{max}=0.56$ at $a=6.3"$, $PA=90^o$). Spiral arms are responsible
for features in the profile at larger distances.

%66
%Dev 87
{\bf NGC 4900} (SBc, $1''\sim 64$ pc, $I=5.3^o$):\\
The galaxy is seen approximately face-on ($I=5.3^o$) and
no twist is found inside the large-scale bar ($e_{max}=0.63$ at
$a=4.9"$, $PA = 140^o$). Further out, an irregular spiral
structure causes changes in the ellipticity and PA profiles.

%67
% Dev87
{\bf NGC 4902} (SBb, $1"\sim 170$ pc, $I=25.1^o$):\\
The PA inside the large-scale bar ($e_{max}=0.60$ at
$a=25"$) is approximately constant (within $10^o$).

%70
% Telesco93
{\bf NGC 5236 (M 83)} (SABc, $I=21.0^o$):\\
The nuclear structure of this nearby ``hot-spot'' starburst galaxy
(e.g. Telesco et al. 1993) is complex also in our JHK images and
the ellipse fitting in the central region not too meaningful
(it even fails in J).
The structure, patchy but dissimilar from optical and mid-IR
will be analyzed elsewhere.
% structure complicated, ell. fitting not aapropriate; see ref.

% 8. NO TRIAXIALITY
% W95
{\bf NGC 6753} (RSAb, $1"\sim 190$ pc, $I=29.3$):\\
The ellipticity and PA profiles of this galaxy
show several small peaks related probably to the presence of
star forming regions and flocculent spiral structure
of the outer disk, as W95 have already noted.
No clear triaxial feature can be found.
% (look to Buta 84: inner ring)
% are the 2-arm sa's dust lanes ? But classified as RSA(r)b, ie no bar

%76 NO TWIST
{\bf IC 1953} (SBd, $1"\sim 110$ pc, $I=44.0^o$):\\
The PA is constant inside the large-scale bar
($e_{max}=0.70$  at $a=23"$, $PA=156^o$).
\vskip2mm

%------------------------------------------------------------------
\begin{table}
\caption{Projected parameters of single bars (H band)} \label{tbl-1}
\vskip-5mm
\begin{center}\scriptsize
\begin{tabular}{llrrrrrrr}
Galaxy&$e_{max}$&$l_{max}$&PA&twist&tw. scale \\
      &         & ('')&($^o$)&($^o$)&('')         \\
%\multicolumn{1}{c}{$P$\tablenotemark{a}} & $P R_{maj}$ & $P R_{min}$ &
%\multicolumn{1}{c}{$\Theta$\tablenotemark{b}} \\
%\tableline
     &    \\
N 1187    & 0.59 & 32  & 132 &  14 & 3-7    \\
N 1255    & 0.53 & 4.4 & 116 &  no & -      \\
N 1302    & 0.35 & 29  & 170 & 20  & 7-29  \\
N 1640    & 0.62 & 31  & 45  & no  & -     \\
N 1784    & 0.68 & 26  & 92  & no  & -     \\
N 1792    & 0.45 & 5   &     & no  & - \\
N 1832    & 0.58 & 17  &166  & 14  & 3-11 \\
%N 2442    & 0.43?& 8?  &52   & 29  & 3-8  \\
N 2525    & 0.69 & 22  &74   & 13  & 3-22 \\
N 3166    & 0.16 & 18  & 170 & 47  &13-17 \\
N 3346    & 0.69 & 14  & 97  & no  & - \\
N 3637    & 0.38 & 15  & 37  & 71  &3-15  \\
N 3673    & 0.73 & 42  & 87  & 17  &3-42  \\
N 3887    & 0.68 & 33  & 0   & 22  &3-25  \\
N 4050    & 0.70?& 69? & 79  & no  & -    \\
N 4106    & 0.37 & 5.3 & 93  & no  & -  \\
N 4267    & 0.22 & 18  & 32  & no  & -  \\
N 4424    & 0.78 & 8   & 110 & no  & - \\
N 4454    & 0.53 & 31  & 22  & 52  & 3-12 \\
N 4519    & 0.58 & 3   & 78  & no  & - & \\
N 4612    & 0.21 & 16  & 98  & 46  & 3-16 \\
N 4665    & 0.53 & 42  & 3   & 11  & 3-42 \\
N 4781    & 0.56 & 6.3 & 90  & no & - \\
N 4900    & 0.63 & 4.9 & 140 & no & - \\
N 4902    & 0.60 & 25  & 67  & no & - \\
N 5643    &0.68?&49? & 87  & 44 & 3-30 \\
N 5701    & 0.44 & 38  & 0   & 39 & 3-39 \\
IC 1953   & 0.70 & 23  & 156 & no & -
\end{tabular}
\end{center}
\end{table}
%------------------------------------------------------------------

%\vfill\eject
%\command \subsection causes error !
%\subsection{Highly inclined galaxies}
\vskip3mm
{\noindent \it 4.5. Highly inclined galaxies}
\vskip3mm

Galaxies listed below have the inclination higher than
$75^o$. We publish their contour plots and
ellipse fitting profiles but do not attempt to interpret them:
\vskip3mm
{\bf NGC 1518} (SBdm, $1"\sim 49$ pc, $I=79.8^o$):
{\bf NGC 2811} (SBa, $1"\sim 152$ pc, $I=85.9^o$),
{\bf NGC 3384} (SB0-/AGN?, $1"\sim 48$ pc, $I=90^o$),
{\bf NGC 3593} (SA0/a, $1"\sim 42$ pc, $I=76.4^o$),
{\bf NGC 3885} (SA0/a, $1"\sim 105$ pc, $I=90^o$),
{\bf NGC 4178} (SBdm, $I=90.0^o$),
{\bf NGC 4192} (SABab/Seyfert 3, $I=90.0^o$),
{\bf NGC 4216} (SABb, $I=90.0^o$),
{\bf NGC 4442} (SB0, $1"\sim 40$ pc, $I=90.0^o$),
{\bf NGC 4461} (SB0+, $1"\sim 120$ pc, $I=85.7^o$),
{\bf NGC 4503} (SB0-, $1"\sim 90$ pc, $I=90.0^o$),
{\bf NGC 4546} (SB0-, $1"\sim 67$ pc, $I=90.0^o$),
{\bf NGC 4684} (SB0+, $1"\sim 100$ pc, $I=90.0^o$),
{\bf NGC 4694} (SB0/HII, $1"\sim 79$ pc, $I=79.7^o$),
{\bf NGC 4856} (SB0/a, $1"\sim 82$ pc, $I=90.0^o$),
{\bf NGC 6810} (SAab, $1"\sim 110$ pc, $I=86.2^o$).

\section{Conclusions}

A sub-sample of 56 galaxies whose nuclear structures
were interpreted in terms of the ellipse fitting on near-IR
images was constructed (16 other galaxies of the survey
are seen nearly edge-on and were
not analyzed in detail). Our principal results are the following.\vskip1mm

1. We classify 17 galaxies of the subsample as having two triaxial structures,
either double bars or a bar with an embedded triaxial bulge (some cases
are speculative because of low-central resolution or high galactic
inclination).
Two of them (NGC 1433 and 6782) were known to possess a double-barred
morphology from previous surveys in other colors, 15 detections are
new (NGC 613, 1079, 1353, 1365, 1398, 1808, 1512, 2217, 2935, 3368, 3393,
4984, 5101, 5566 and ESO 437-67).
\vskip1mm

2. We find 16 other nuclear isophotal twists
not associated with a clear ellipticity maximum (14 of them are new:
NGC 1187, 1832, 2442, 2525, 2997, 3166, 3637, 3673, 3887, 4454,
4612, 4665, 5643, 5701; twists in NGC 1302 and 1512 were known from other
bands).
\vskip1mm

3. We detect central triaxial features (nuclear bars or triaxial bulges)
in three galaxies classified as SA in RC3: NGC 1792, 4438 and 5427.
On the other hand, we find no evidence for triaxiality in the following
SA's: NGC 2911, 4212, 4501, 4689 and 6753.
\vskip1mm

4. Four of our 17 double-barred galaxies host a Seyfert nucleus:
NGC 613, 1365, 1808, 3393 (Seyfert activity of NGC 1808 is controversial).

\vskip1mm

5. Among 9 Seyferts from the sub-sample, there are 4 with
a double bar (see item 4), 2 with a nuclear bar (NGC 4438 and 5427),
1 with a strong twist along its large-scale bar (NGC 5643) and
2 show no clear twist (NGC 2911 and 4501).
\vskip1mm

6. Nine of our double-barred galaxies are known to have a nuclear ring
(NGC 1433, 1512, 1808, 2935, 4984, 6782 and ESO 437-67) or a nuclear spiral
(NGC 613 and 1365).
\vskip1mm

7. Among 32 galaxies with double barred structure or twist,
29 are of early Hubble types (S0-Sc);
only 3 are of late type Sc: NGC 1187, 2525, 2997, 5643.
Twists in the first 3 Sc's are small ($10-20^o$) and can be caused
by projection effects; the twist in NGC 5643 -- which is a Seyfert emitter
-- is strong and probably intrinsic. No Sc galaxy with a twist has
been known so far.
\vskip1mm

Comparison of the present survey with photometry in other
colors and with kinematical data, as well as the interpretation,
will be done in a separate paper.

\begin{acknowledgements}
The authors acknowledge the participation of M. Shaw on the
selection of the observed sample of galaxies.
B.J. wishes to express many thanks to P. Lena,
J.-P. Zahn and J. Palou\v s for their efforts in co-organizing the
PhD ``cotutelle'' between
University Paris VII and Charles University in Prague, and, especially,
to F. Combes for having accepted to
co-direct his PhD thesis.
He also acknowledges useful discussions on data reduction techniques
with J.M. Deltorn as well as the help of H. Flores with software problems.
The stay of B.J. at the Paris Observatory was supported by a
scholarship of the French Go\-vern\-ment.
\end{acknowledgements}

\vskip0.8cm

\noindent
{\bf Appendix}

\vskip-2mm
\begin{figure} [hbp] 
\caption[]{
Contour plots and ellipse fitting. The left box
shows logarithmically scaled intensity map on the area of 120 x 120
arcseconds; the North is at the top, the East at the left.
The ellipse fitting output -- radial profiles
of ellipticity, position angle and surface brightness --
is displayed in the central plot; the abscissa -- semi-major axis of
fitted ellipses -- is scaled logarithmically. The right plot
shows the same quantities after deprojection. The vertical dash-dot
lines in radial profile plots separate the innermost region
of $a<3''$ where the ellipse fitting is judged to be unreliable.}
\end{figure}

\end{document}